\documentclass[a4paper,11pt]{article}
\usepackage{jheppub} 
\usepackage{lineno}
\usepackage{color,soul}
\usepackage{microtype}
\usepackage[english]{babel}
\usepackage[utf8]{inputenc}
\usepackage{amssymb}
\usepackage{mathtools,amsmath}
\usepackage{slashed}
\usepackage{physics}
\usepackage{breqn}
\usepackage{cancel}
\usepackage{comment}
\usepackage[dvipsnames]{xcolor}
\usepackage{listings}
\usepackage{hyperref}
\usepackage{mathbbol}
\usepackage{amsthm}
\usepackage{empheq}
\usepackage[symbol]{footmisc}

\newcommand{\be}{\begin{equation}}
\newcommand{\ee}{\end{equation}}
\newcommand{\bee}{\begin{equation*}}
\newcommand{\eee}{\end{equation*}}

\newcommand{\ms}[1]{}
\newcommand{\dm}[1]{}
\newcommand{\as}[1]{}
\newcommand{\YB}[1]{}
\newcommand{\mg}[1]{}
\newcommand{\mz}[1]{}

\newcommand{\fig}[1]{Fig.~\ref{#1}}

\arxivnumber{2412.xxxxx} 

\title{\boldmath Bubble dynamics in a QCD-like phase diagram}

\subheader{\begin{flushright}
\texttt{CPHT-RR089.122024}
\end{flushright}}

\author[a]{Yago~Bea,}
\author[b]{Mauro~Giliberti,}
\author[a,c]{David~Mateos,}
\author[d]{Mikel~Sanchez-Garitaonandia,}
\author[e]{Alexandre~Serantes}
\author[f]{and Miguel~Zilh\~ao}

\affiliation[a]{Departament de F\'\i sica Qu\`antica i Astrof\'\i sica and Institut de Ci\`encies del Cosmos (ICC),  \\ Universitat de Barcelona, Mart\'\i\  i Franqu\`es 1, ES-08028, Barcelona, Spain.}
\affiliation[b]{Dipartimento di Fisica e Astronomia, Università degli Studi di Firenze; Via G. Sansone 1; I-50019 Sesto Fiorentino (Firenze), Italy.}
\affiliation[c]{Instituci\'o Catalana de Recerca i Estudis Avan\c cats (ICREA), Passeig Llu\'\i s Companys 23,  ES-08010, Barcelona, Spain.}
\affiliation[d]{CPHT, CNRS, \'Ecole polytechnique, Institut Polytechnique de Paris, 91120 Palaiseau, France.}
\affiliation[e]{Department of Physics and Astronomy, Ghent University, 9000 Ghent, Belgium.}
\affiliation[f]{Centre for Research and Development in Mathematics and Applications (CIDMA),\\ Department of Mathematics, University of Aveiro, 3810-193 Aveiro, Portugal.}

\emailAdd{yagobea@icc.ub.edu}
\emailAdd{mauro.giliberti@unifi.it}
\emailAdd{dmateos@fqa.ub.edu}
\emailAdd{mikel.sanchez@polytechnique.edu}
\emailAdd{alexandre.serantesrubianes@ugent.be}
\emailAdd{mzilhao@ua.pt}

\abstract{A line of first-order phase transitions is conjectured in the phase diagram of Quantum Chromodynamics at non-zero baryon density.

If this is the case, numerical simulations of neutron star mergers suggest that various regions of the stars may cross this line multiple times. This results in the nucleation of bubbles of the preferred phase, which subsequently expand and collide. The resulting gravitational wave spectrum is highly sensitively to the velocity of the bubble walls. We use holography to perform the first microscopic simulation of bubble dynamics in a theory that qualitatively mirrors the expected phase diagram of Quantum Chromodynamics. We determine the wall velocity in the metastable regions and we compare it to theoretical estimates. We discuss implications for  gravitational wave  production. 
}

\begin{document}
\maketitle
\flushbottom

\section{Introduction}\label{sec:intro}
\begin{figure}[t]
    \begin{center}
    	\includegraphics[width=0.99\textwidth]{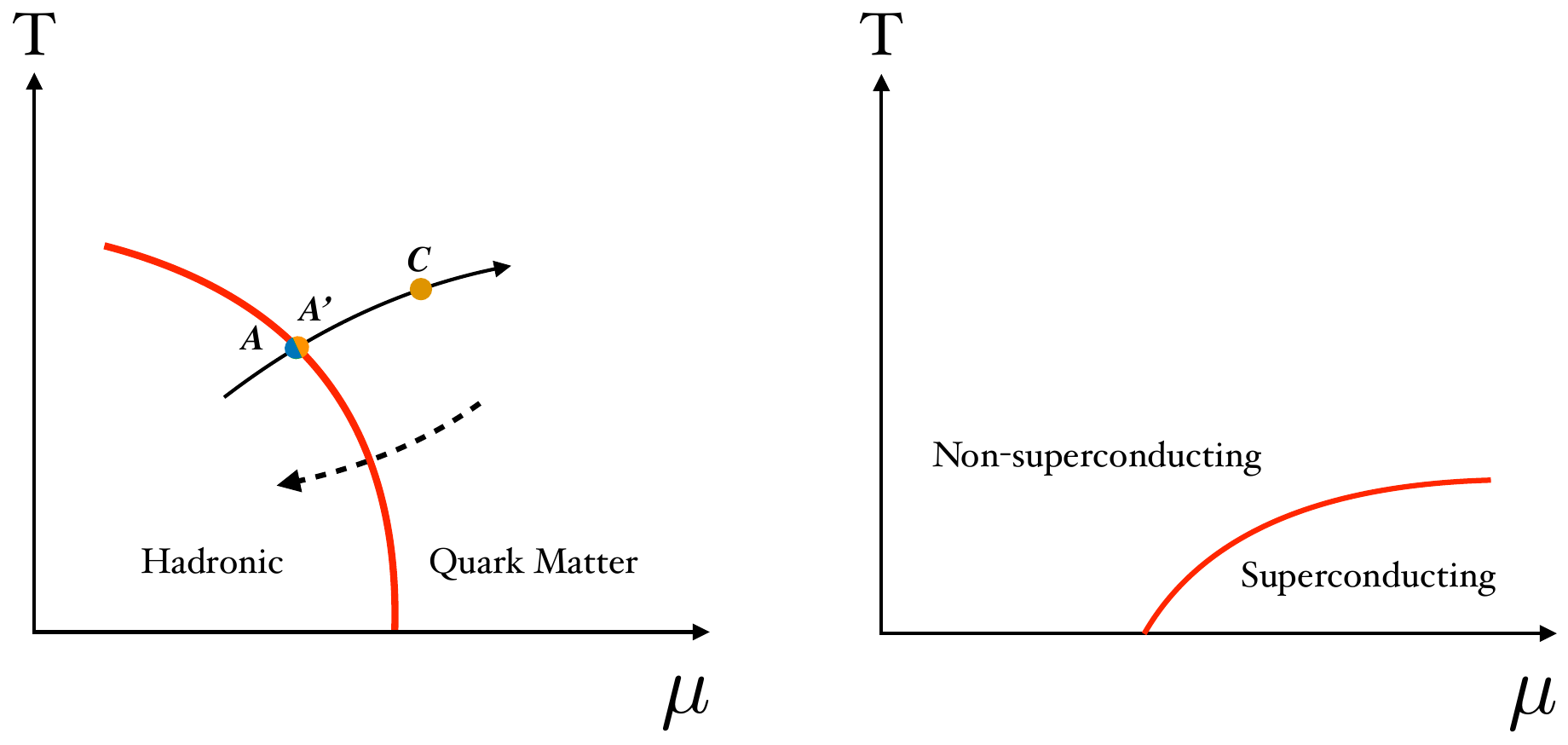}
    \end{center}
    \caption{\small Two possible phase transitions in QCD, indicated by the solid red curves.  $T$ and $\mu$ are the temperature and the baryon chemical potential, respectively.  The dotted (solid) black curve on the left panel shows a possible evolution of a region of a NS merger as this region is supercooled (superheated). 
    The points dubbed $A$, $A'$ and $C$ correspond to the states shown in \fig{meta}. See text and Ref.~\cite{Casalderrey-Solana:2022rrn}.}
\label{FOPTquark}
\end{figure}

A  variety of arguments suggest that at least two first-order phase transitions (FOPT) may be present in the phase diagram of Quantum Chromodynamics (QCD) as a function of temperature and baryon chemical potential  \cite{Stephanov:2004wx,Alford:2007xm,Fukushima:2010bq,Guenther:2022wcr}, as sketched in \fig{FOPTquark}. One is the transition from hadronic matter to quark matter. The other is the transition from a non-superconducting 
to a color-superconducting phase. While these transitions are well motivated, rigorously establishing their existence is a fundamental open problem in nuclear and particle physics whose solution has resisted both theoretical and experimental attempts for decades.

Gravitational waves produced in neutron star (NS) mergers could  provide direct experimental access to these phase transitions \cite{Casalderrey-Solana:2022rrn}. Numerical simulations of NS mergers based on equations of state (EoS) with a hadronic-quark matter phase transition include \cite{Most:2018eaw,Most:2019onn,Ecker:2019xrw,Prakash:2021wpz,Weih:2019xvw,Tootle:2022pvd,Fujimoto:2022xhv}. These studies have shown that the dynamics of the merger results in the formation of regions in which the matter is sufficiently heated and/or compressed  that the thermodynamically preferred phase is the quark-matter phase. In other words, the matter in these regions is pushed along the solid black curve in \fig{FOPTquark}. For brevity, we will refer to these regions simply as ``superheated regions''. When this matter cools down again, the phase transition is crossed in the opposite direction, as indicated by the dotted black curve in \fig{FOPTquark}. We will refer to the resulting regions as ``supercooled regions''. 
Although no simulation based on an EoS with a color-superconducting phase has been performed, the large baryon densities found in existing simulations make it conceivable that color-superconducting matter may also be formed in NS mergers.

\begin{figure}[tbp]
\begin{center}
    \includegraphics[width=0.75\textwidth]{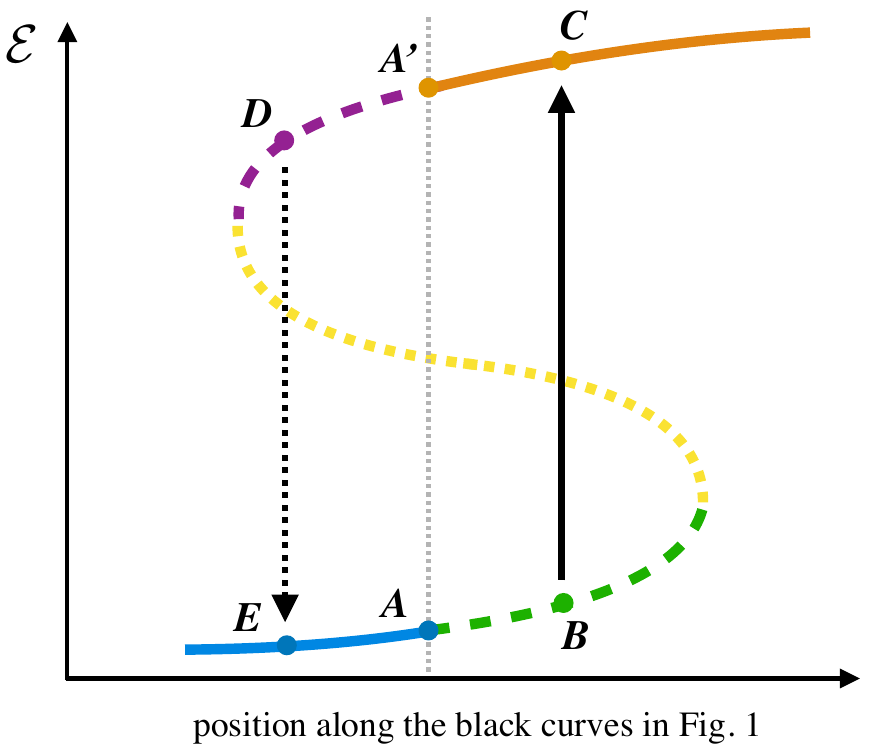}
  \caption{\small Energy density as a function of the position along the  black solid curve in \fig{FOPTquark}. Both $T$ and $\mu$ increase from left to right. The grey, dotted,  vertical line indicates the location of the phase transition, 
   determined by the condition that the states $A$ and $A'$ have the same free energy. The orange and blue curves represent stable states, the purple and green curves denote metastable states, and yellow curve indicates unstable states. As some region of the NS merger is sufficiently  heated and/or compressed,  it enters the lower metastable branch. At the point $B$ bubbles of the preferred state $C$ on the upper stable branch are  nucleated. The direction of this phase transition is the opposite of that in a supercooled phase transition, which takes place from $D$ to $E$, as indicated by the vertical, dotted, black arrow.
   }
   \label{meta}
\end{center}
\end{figure}

The energy density along the black solid curve in \fig{FOPTquark} displays the multivalued form characteristic of a FOPT, as  shown in \fig{meta}. The superheated region is pushed from left to right along the lower branch of the phase diagram. Once the superheating is large enough, namely once the region of interest is sufficiently deep into the metastable branch, bubbles of the stable phase begin to nucleate. The point where this happens is labeled ``$B$'' in  \fig{meta}, the nucleated phase is labeled ``$C$'', and the transition is indicated by a vertical, solid, black arrow.  We will refer to these nucleated bubbles as ``superheated bubbles''. The direction of the ``superheated'' transition is opposite to that of a cosmological FOPT. In the latter, the Universe is supercooled, namely pushed from right to left along the upper branch into the upper metastable region, until it transitions down to the lower stable branch, as indicated in \fig{meta} by the dotted, vertical, black arrow. We will refer to the nucleated bubbles in this case as ``supercooled bubbles''. This type of transition will also occur in a NS merger when the superheated region cools down again.

The bubble dynamics in a NS merger can produce gravitational waves that would provide direct experimental access to the QCD phase transition \cite{Casalderrey-Solana:2022rrn}. The order-of-magnitude estimate in this reference shows that the frequency of these gravitational waves falls roughly in the MHz range, and that they may be potentially observable with future detectors --- see e.g.~\cite{Berlin:2023grv}. 

Refining the analysis of Ref.~\cite{Casalderrey-Solana:2022rrn} requires a precise description of the bubble dynamics. Once the bubbles have grown to a macroscopic size, this is provided by hydrodynamics. The hydrodynamics of relativistic, supercooled bubbles have been extensively studied in the literature, see e.g.~\cite{Espinosa:2010hh}. The case of superheated bubbles has been recently addressed~\cite{Barni:2024lkj,Bea:2024bxu}. Like any effective theory, hydrodynamics requires input from the microscopic theory, specifically the equation of state  (EoS) and the bubble wall velocity. The latter depends on out-of-equilibrium physics and its computation  is challenging even in weakly-coupled theories
 \cite{Moore:1995si, Dorsch:2018pat, Lewicki:2021pgr, Laurent:2022jrs, Jiang:2022btc}. Nevertheless, determining this velocity is crucial because the gravitational wave spectrum is highly sensitive to it \cite{Espinosa:2010hh, Caprini:2019egz}. While certain approximation schemes exist \cite{Ai:2023see, BarrosoMancha:2020fay, Ai:2021kak, Janik, Sanchez-Garitaonandia:2023zqz, Ai:2024uyw,Ekstedt:2024fyq}, it is interesting to study it in systems where a first-principle calculation is possible. 
 In this paper, we will use holography to perform the first microscopic calculation of the wall velocity  in a model with a phase diagram like that in \fig{FOPTquark}(left). We thus extend the original holographic calculation \cite{Bea:2021zsu} in two significant ways: first, by considering superheated bubbles, and second, by incorporating a non-zero baryon chemical potential.

\section{Holographic model}\label{sec:model}
Following \cite{DeWolfe:2010he}, in order to include a conserved $U(1)$ current in the boundary theory, we add a Maxwell field to the five-dimensional Einstein-scalar model of \cite{Bea:2018whf}. The action takes the form 
\be\label{eq:action}
    S = \frac{2}{\kappa_5^2}\int d^5x\sqrt{-g}\left(\frac{1}{4}\mathcal{R}-\frac{1}{2}(\partial\phi)^2-V(\phi)-\frac{f(\phi)}{8}F^2\right),
\ee
where $\kappa_5$ is the five-dimensional gravitational coupling constant. We choose units such that  $\kappa_5 = 8\pi$. 

The $U(1)$ gauge symmetry associated to the Maxwell field in the bulk is dual to a global $U(1)$ symmetry in the boundary theory that we will think of as baryon number. The scalar potential $V(\phi)$ encodes the thermodynamics of the boundary theory at zero baryon number, but it also affects the properties at non-zero baryon number. The function $f(\phi)$ that sets the normalization of the kinetic term for the gauge field affects only the thermodynamics at non-zero baryon number. The key point is that $V(\phi)$ and $f(\phi)$ can be chosen so that the phase diagram of the boundary theory mirrors that in \fig{FOPTquark}(left) \cite{DeWolfe:2010he}.

The  equations of motion derived from the action \eqref{eq:action} read
\be\label{eq:eoms}
\begin{aligned}
    R_{\mu\nu} - \frac{1}{2}g_{\mu\nu} - 8\pi T_{\mu\nu} & = 0 \,,\\
    \Box \phi -\partial_{\phi} V(\phi) - \frac{\partial_{\phi} f(\phi)}{8}F^2 & = 0\,,\\
    \nabla_{\mu}\left(f(\phi) F^{\mu\nu}\right) & = 0\,,
\end{aligned}
\ee
with
\be\label{eq:Bulk_Tmunu}
8\pi T_{\mu\nu} = 2\partial_{\mu}\phi\partial_{\nu}\phi + f(\phi)F_{\mu\alpha}F_{\nu}^{\alpha} -g_{\mu\nu}\left(\partial_{\alpha}\phi\partial^{\alpha}\phi+2V(\phi)+\frac{f(\phi)}{4}F^2\right).
\ee
As in \cite{Bea:2018whf}, we consider a scalar potential that can be derived  from the superpotential $W(\phi)$ 
\be
    LW(\phi) = -\frac{3}{2} -\frac{\phi^2}{2} + \lambda_4\phi^4 + \lambda_6\phi^6,
    \label{eq:superpotential}
\ee
via the standard relation 
\be
    V = - \frac{4}{3}W^2 + \frac{1}{2}\left( \partial_\phi W \right)^{2}.
\ee
This potential has a maximum at 
$\phi = 0$, corresponding to an ultraviolet (UV) fixed point of the dual gauge theory. The mass of the scalar field around this maximum, $m^2 = -3/L^2$, implies that the scalar field is dual to an scalar operator of scaling dimension $\Delta_{UV} = 3$ with respect to the UV fixed point. The boundary theory is therefore a conformal field theory (CFT) deformed by a dimension-three, relevant operator. The source for this operator, $\Lambda$, breaks conformal invariance and sets the characteristic, intrinsic scale in the boundary theory. We will use the values\footnote{In the terminology of \cite{Bea:2018whf} this correspond to $\phi_M=1.1,~\phi_Q=10$.}
\begin{equation}
\label{lambdachoice}
\lambda_4 = -0.206612 \,,\qquad 
\lambda_6 = 0.1 \,.
\end{equation}
Under this assumption, the potential has a minimum at a positive value of $\phi$, which corresponds to an infrared (IR) fixed point of the dual gauge theory. 

For vanishing Maxwell field, the equilibrium and non-equilibrium properties of this family of models were extensively studied in \cite{Attems:2016ugt, Attems:2016tby, Attems:2017ezz, neutral, Ares:2020lbt, bubblevelocity,Bea:2021ieq, Bea:2021zol, Jecco}. In particular, for the choice \eqref{lambdachoice} the theory exhibits a smooth crossover as a function of temperature at zero chemical potential. 
For appropriate choices of $f(\phi)$, the crossover turns into a line of FOPTs at non-zero chemical potential. We will work with the simple choice  
\be
f(\phi) = \frac{e ^{-3\phi/4}}{32} \,.
\ee

The functions $V(\phi)$ and  $f(\phi)$ completely specify the model. Unlike in some previous work \cite{DeWolfe:2010he,Critelli:2017oub,Grefa:2021qvt,Tootle:2022pvd,Hippert:2023bel,Shah:2024img}, the motivation for our choice is not to match the known features of QCD (for example, the EoS at zero chemical potential) as precisely as possible at a quantitative level. Instead, we wish to provide a first study of the out-of-equilibrium bubble dynamics in a phase diagram that mirrors that of QCD qualitatively can be studied holographically. For this purpose, we have chosen a model that makes the numerical analysis as tractable as possible.  
In particular, this will allow us to  determine the bubble wall velocity almost everywhere in the phase diagram. This includes points of low temperature and high chemical potential with $T/\mu \simeq 0.03$ 
where, as we will see, the wall velocity reaches its maximum values. Exploring smaller values of this ratio is challenging  even in our model for  reasons mentioned in Sec.~\ref{sec:phase diagram}. Extending our analysis to other models that incorporate quantitative features of QCD is likely to increase the technical complexity but is conceptually straightforward. We leave this analysis for future work.

\section{Phase diagram}
\label{sec:phase diagram}
To determine the phase diagram, we follow \cite{DeWolfe:2010he} closely,\footnote{One difference with the model used in that reference is that, in our case, the gravity solution at zero temperature and zero chemical potential is regular.} including the choice of coordinates and the choice of gauge for the gauge field, which differ from those used in Sec.~\ref{time}. We construct a large family of black brane solutions. An analysis of the near-horizon behavior shows that these can be parametrized by the value of only two independent boundary conditions at the horizon. We choose these to be the scalar field and  the derivative of the time-component of the Maxwell field with respect to the holographic direction, $r$, and we label them as  
\begin{equation}
    (\phi_H,\Phi_H) \equiv (\phi, \partial_r A_t)|_{r=r_H}.
\end{equation}
Operationally, we construct a  grid in these variables. For each point on it, we  solve the equations and from the solutions we read off the  expectation values of the stress tensor, $T_{\mu\nu}$, and of the baryon current, $j^\mu$. From these we extract the energy density 
$\mathcal{E}$, the pressure $\mathcal{P}$, the baryon chemical potential $\mu$ and the baryon density 
$\rho$. The necessary holographic renormalization procedure is the same as in e.g.~\cite{Attems:2017zam,Casalderrey-Solana:2020vls}. In particular, the presence of the gauge field does not require the addition of any new counterterms \cite{Kim:2016dik}.
The temperature $T$ and the entropy density $s$ are obtained from the near-horizon geometry.  The free energy density is obtained through the usual thermodynamic relations
\mbox{$\mathcal{F}=-\mathcal{P}=\mathcal{E}-T s -\mu \rho$}. 

\begin{figure}[tbp]
\centering
\includegraphics[width=0.85\textwidth]{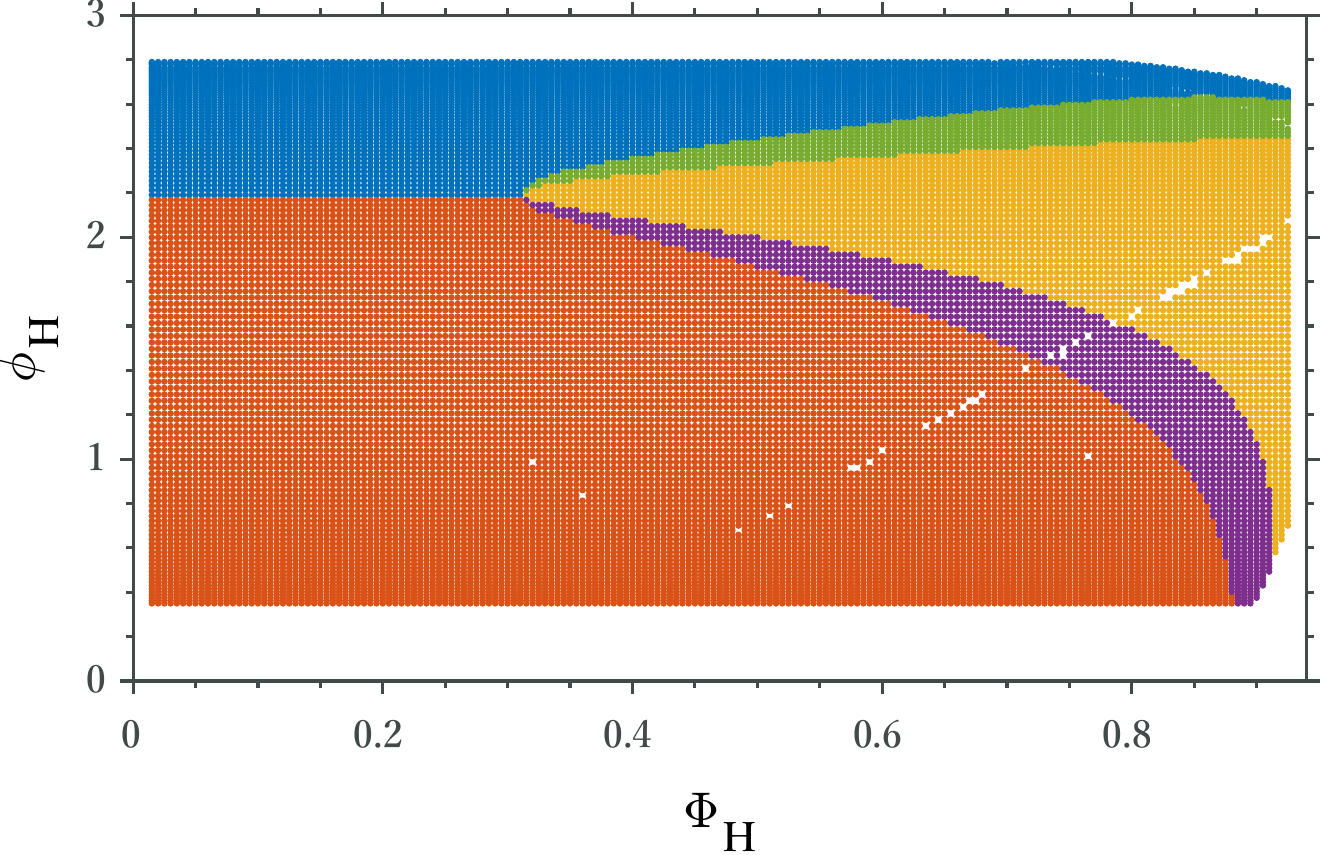}
\caption{Equilibrium solutions labeled by the  boundary conditions at the horizon, $(\phi_H,\Phi_H)$. Orange and blue states are globally stable, purple and green states are metastable, and yellow states are locally unstable. The point where all these colors meet is the critical point.}
\label{fig:all_eq_solutions}
\end{figure}

The set of equilibrium states that we studied is shown in \fig{fig:all_eq_solutions}.
\begin{figure}[tbp]
\centering
\includegraphics[width=.95\textwidth]{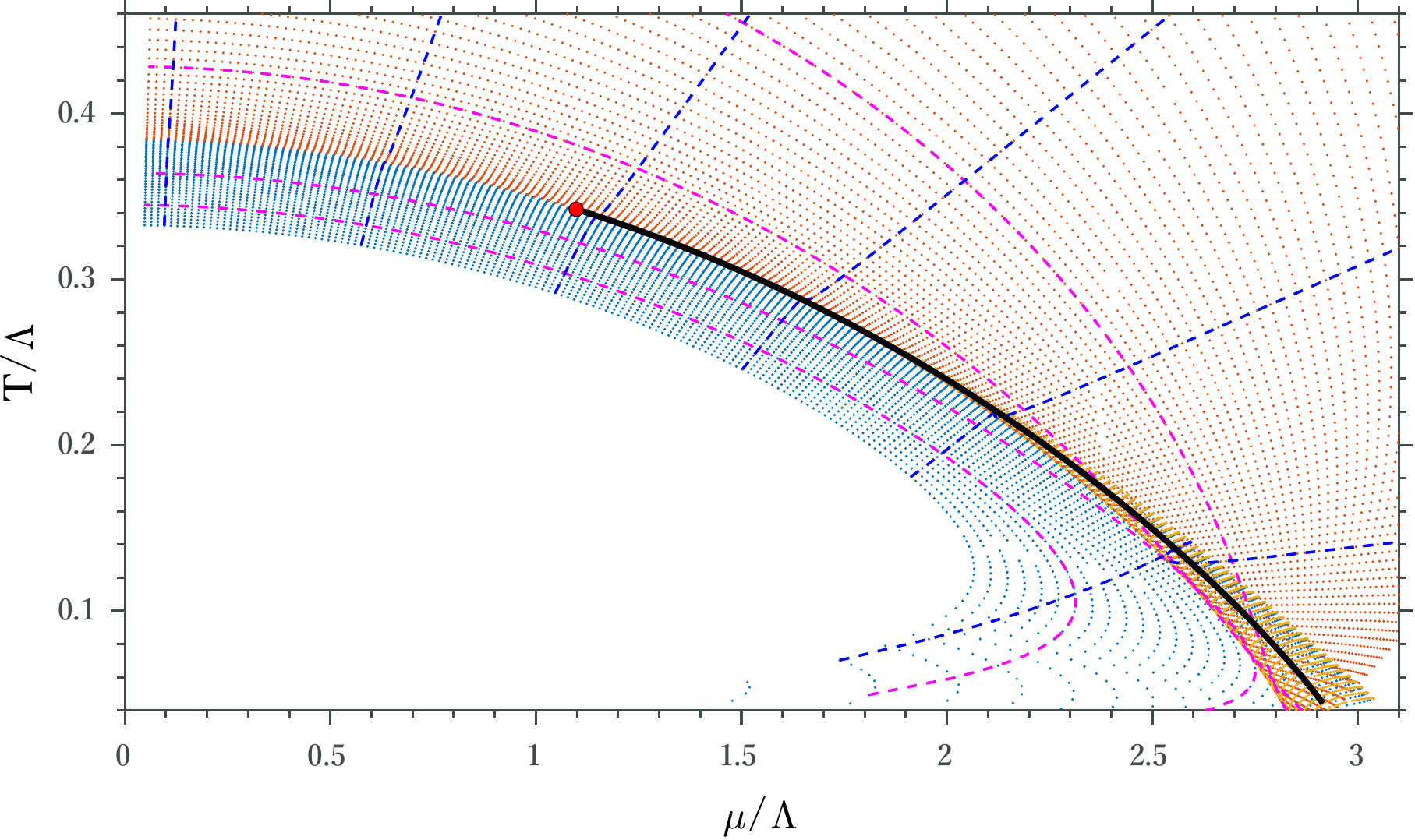}
\caption{Map from  the $(\phi_H,\Phi_H)$ variables to $(T,\mu)$. Each point corresponds to one state. Dashed blue curves correspond to constant values of $\Phi_H$, with $\Phi_H$ increasing from left to right. Dashed pink curves correspond to constant values of $\phi_H$,  with $\phi_H$ decreasing outwards.
The continuous black curve indicates the location of the FOPT, which  terminates at the CP $(T, \mu)= (0.34, 1.09)\Lambda$, marked by a red dot. 
}
\label{tmu_line_crossover}
\end{figure}
The $(\phi_H,\Phi_H)$ variables are in one-to-one correspondence with them. As expected in the presence of a FOPT, $(T,\mu)$ do not share this property. The map from the former to the latter is shown in \fig{tmu_line_crossover}. 
In this figure we observe the backtracking of the constant-$\Phi_H$ curves at high chemical potential. This gives rise to 
the expected multivaluedness associated to a FOPT. In fact, as we will see, this multivalued region coincides exactly coincides with the region where metastable states exist.

\begin{figure}[tbp]
\centering
\includegraphics[width=.95\textwidth]{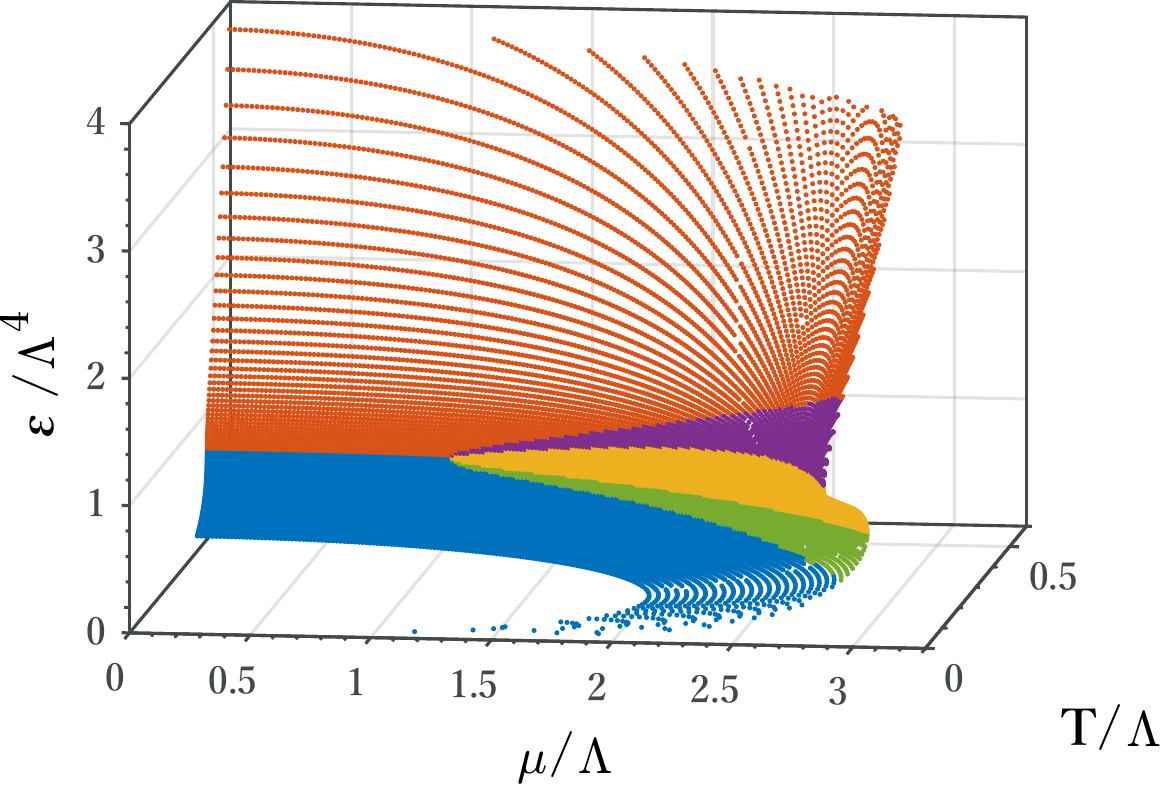}\\[10mm]
\includegraphics[width=.95\textwidth]{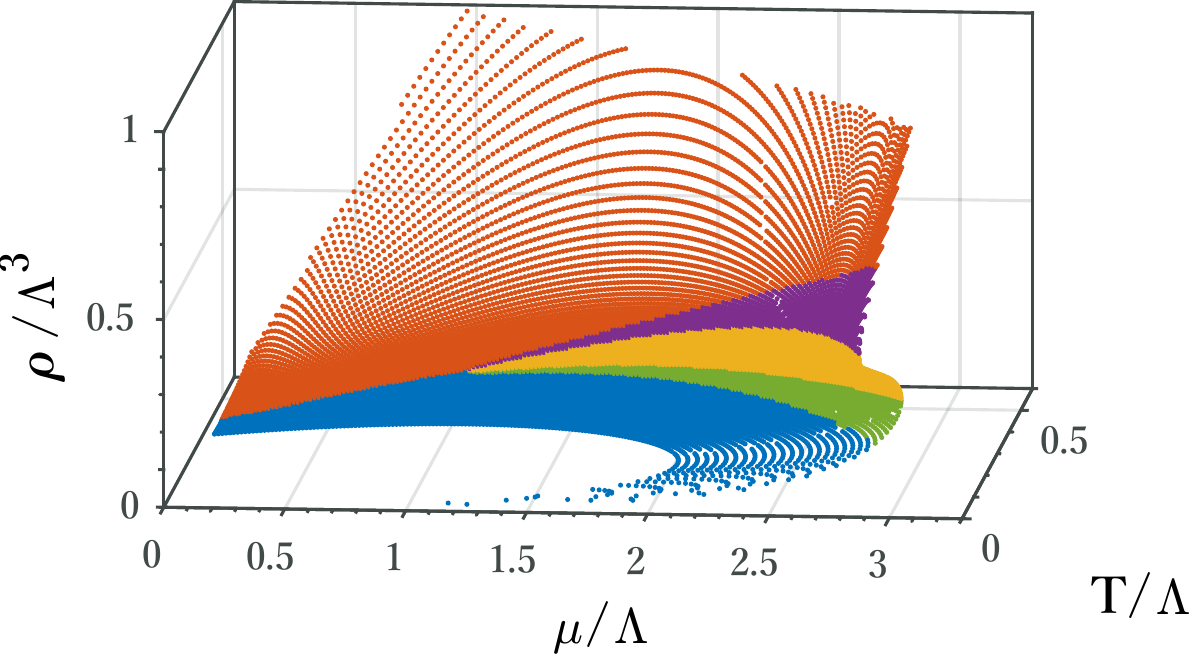}
\caption{Energy (top) and baryon (bottom) densities as functions of temperature and baryon chemical potential.  The color coding is the same as in \fig{fig:all_eq_solutions}.
}
\label{fig:EvsTvsmu}
\end{figure}

\begin{figure}[tbp]
\centering
\includegraphics[width=.99\textwidth]{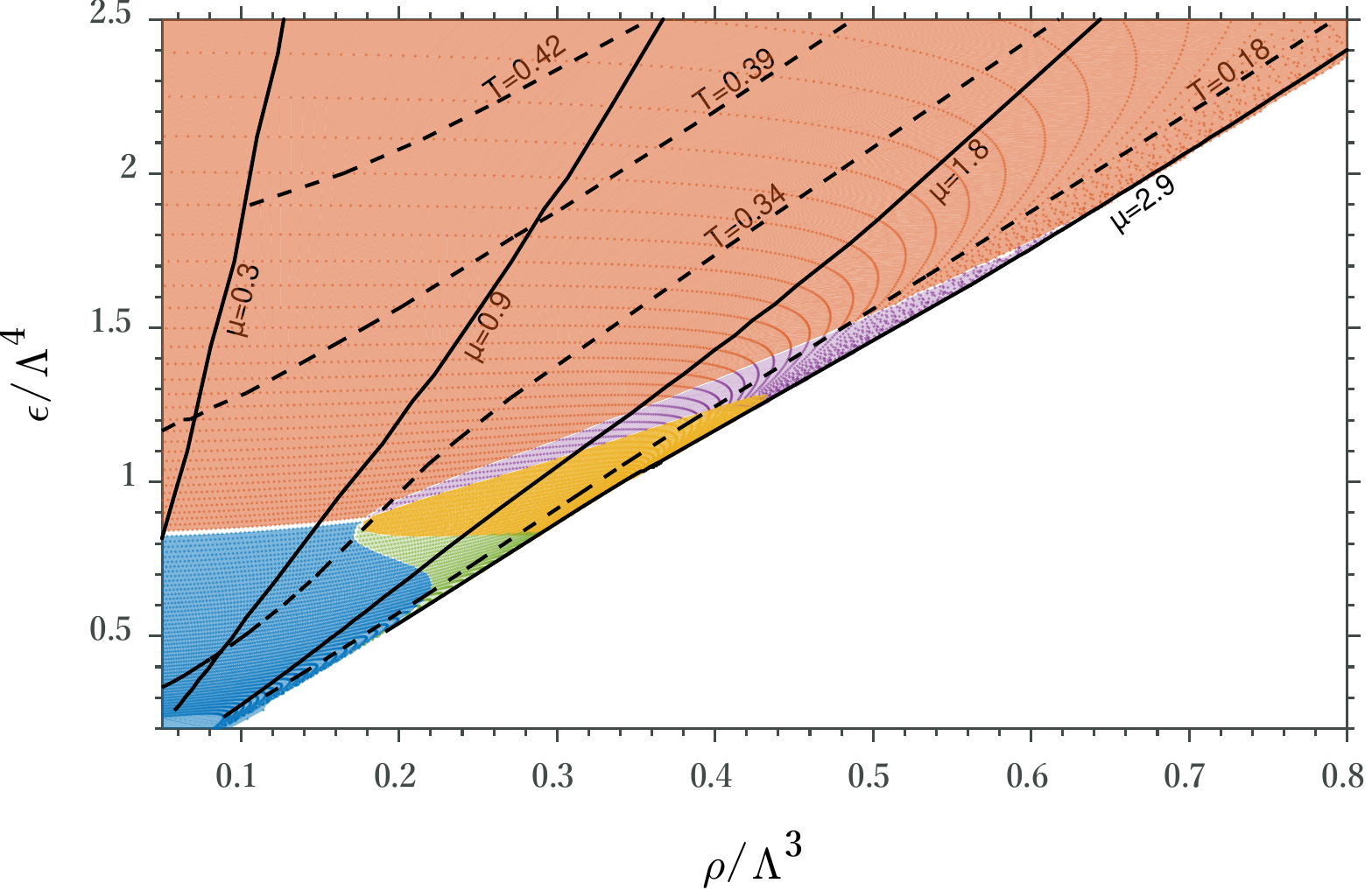}
\caption{Map from the $(T,\mu)$ variables to the $(\mathcal{E},\rho)$ variables, with constant-$T$ and constant-$\mu$ (in units of $\Lambda$) curves shown as black dashed and black solid curves, respectively. 
}
\label{tmuerho}
\end{figure}

The energy and the baryon densities\footnote{Rescaled by a factor of $\kappa_5^2/2L^3$, as in Eq.~\eqref{eq:VEVs}.}  as functions of $(T, \mu)$ are shown in \fig{fig:EvsTvsmu}, and the map  from the $(T,\mu)$ variables to the  $(\mathcal{E},\rho)$ variables is displayed in 
\fig{tmuerho}. 
$(\mathcal{E},\rho)$ are in one-to-one correspondence with equilibrium states across most of the phase diagram except in a small  region near the critical curve at low temperature and high chemical potential. In this region there exists more than one state for a given pair $(\mathcal{E},\rho)$. Qualitatively, this situation is analogous to the energy-multivaluedness in the top plots of Figs.~7 and 8 in Ref.~\cite{Bea:2021zol}. This feature makes the time evolution of bubbles numerically very challenging, so we will avoid this region in this paper.   

\begin{figure}[tbp]
\centering
\includegraphics[width=.95\textwidth]{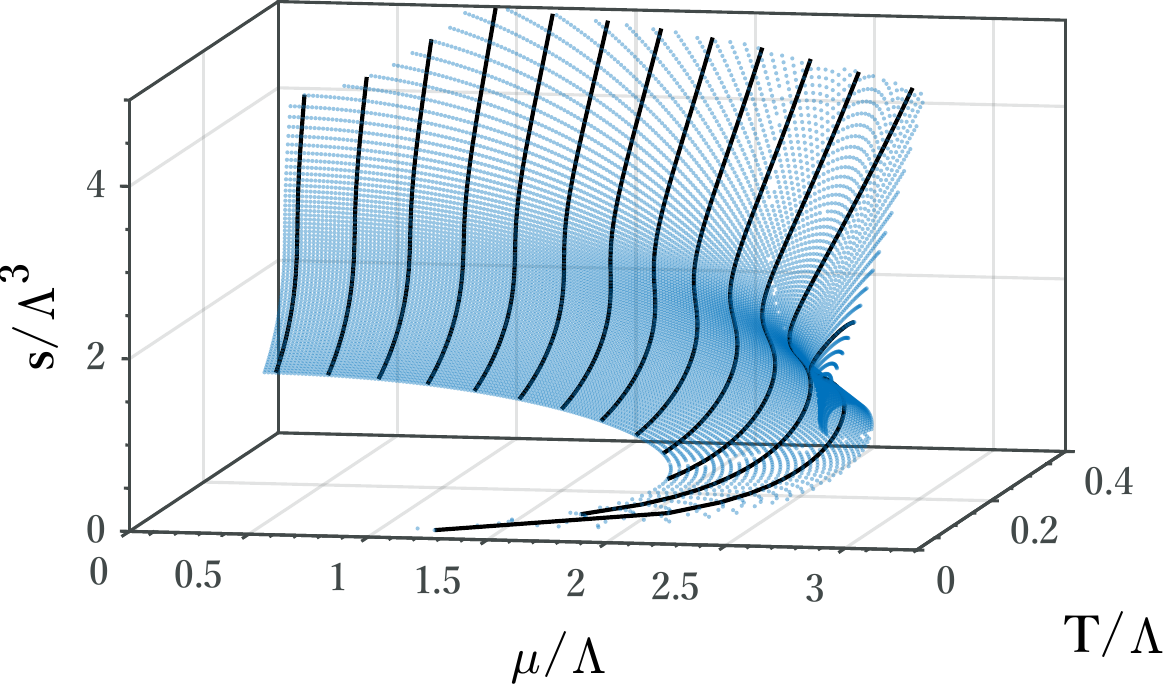}
\caption{Entropy density as a function of temperature and baryon chemical potential.  The black curves of constant $\Phi_H$ are shown to make the multivaluedness clearer.}
\label{entropy}
\end{figure}

\begin{figure}[htbp]
\centering    {\includegraphics[width=.9\textwidth]{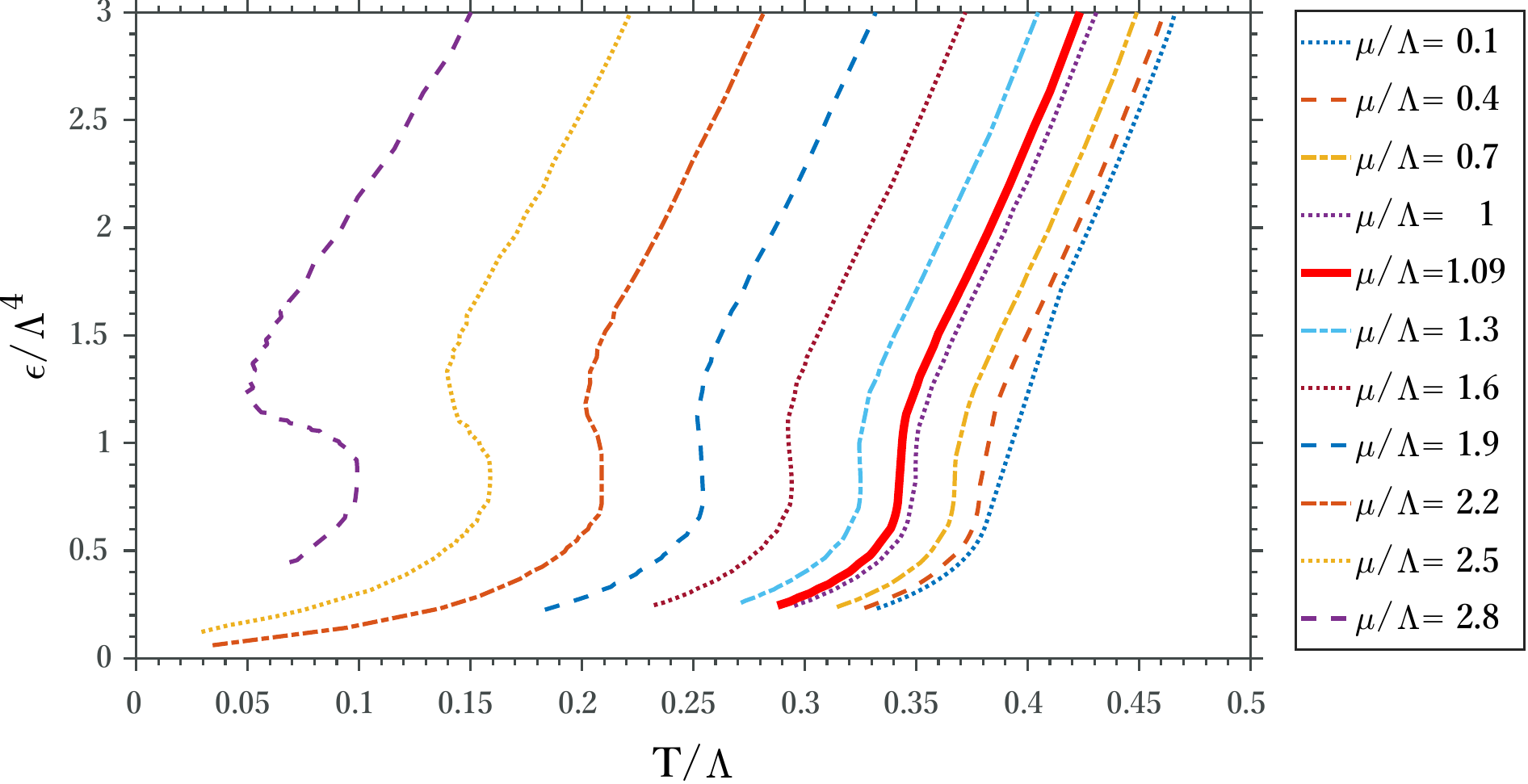}\\
    \vspace{10mm}
    \includegraphics[width=.9\textwidth]{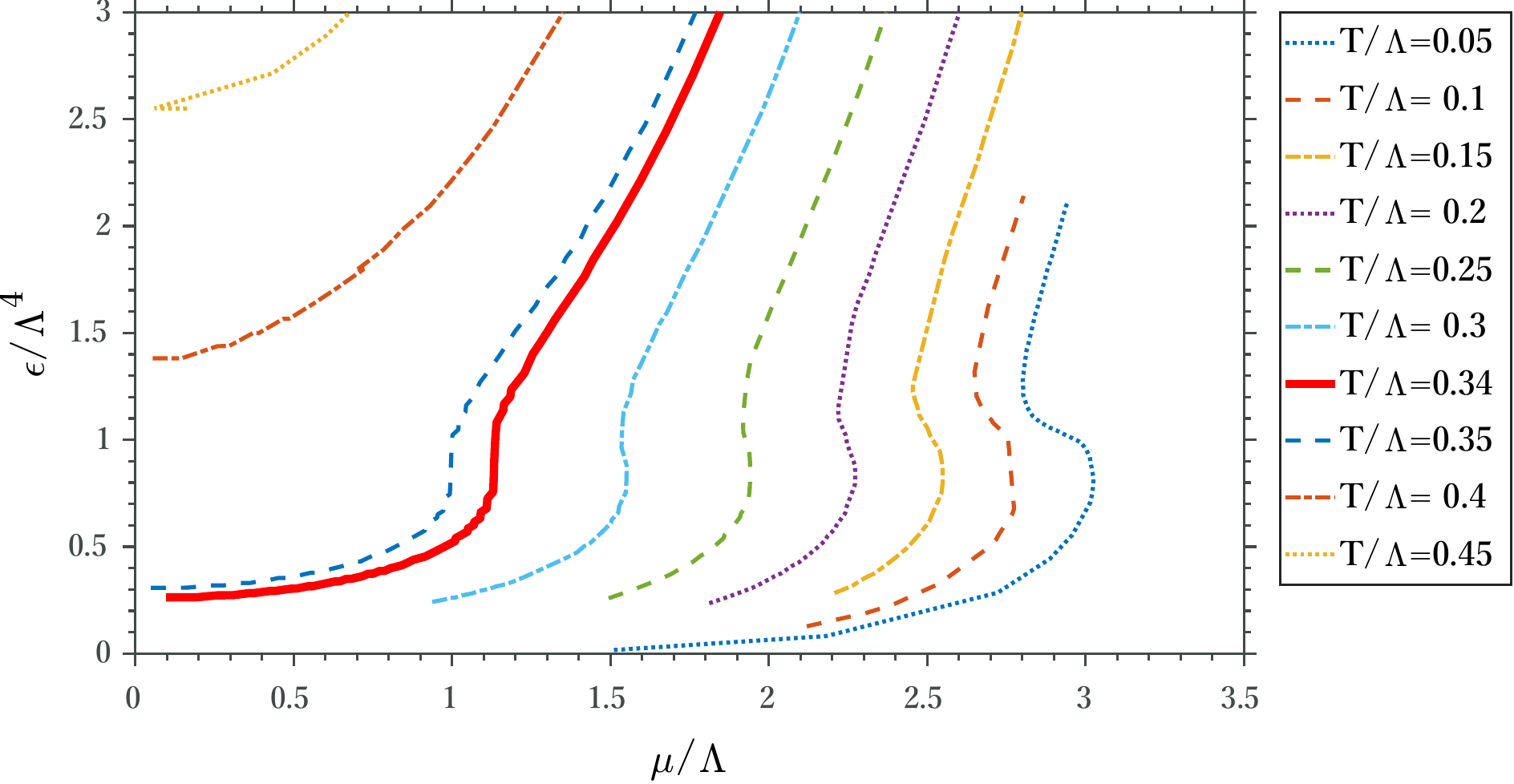}}
\caption{Constant-$\mu$ (top) and constant-$T$ (bottom) slices of the energy density. The thick red curves correspond to a second-order phase transition.}
\label{fig:slices_energy}
\end{figure}

The color coding in Figs.~\ref{fig:all_eq_solutions}, \ref{tmu_line_crossover} and \ref{fig:EvsTvsmu} indicates the stability properties of each state, as we explain below. In particular, the orange and the blue regions correspond to the high- and the low-density phases. The point where all these colored regions meet is the critical point (CP), located at $(T, \mu)= (0.34, 1.09)\Lambda$. To the left of it the transition is a crossover. To the right of the CP we observe the  characteristic multivaluedness of a FOPT. To illustrate this aspect more clearly, in \fig{entropy} we show the entropy density with lines of constant $\Phi_H$ to guide the eye, and in \fig{fig:slices_energy} we show constant-$T$ and constant-$\mu$ slices of the energy density. 
We see that, at low chemical potential, the thermal physics corresponds to a crossover. At $\mu=1.09\Lambda$  this turns into a second-order phase transition, and above this value into a FOPT. In the crossover region there is no sharp distinction between the high- and the low-density  phases. For operational reasons, we have chosen this separatrix to be a line of constant $\phi_H \simeq 2.2$. 

Physically, in this region one may define the location of the crossover in different ways. One natural definition is as the point of maximum slope on a constant-$\mu$ curve, namely, as the point where $(\partial \mathcal{E}/\partial T)_\mu$

is maximized. Selecting such points for different $\mu$'s gives a curve in the phase diagram. We have checked that our operational definition provides an excellent approximation to this curve, to within the size of a few points in \fig{tmu_line_crossover}.

\begin{figure}[htbp]
\centering
\includegraphics[width=0.95\textwidth]{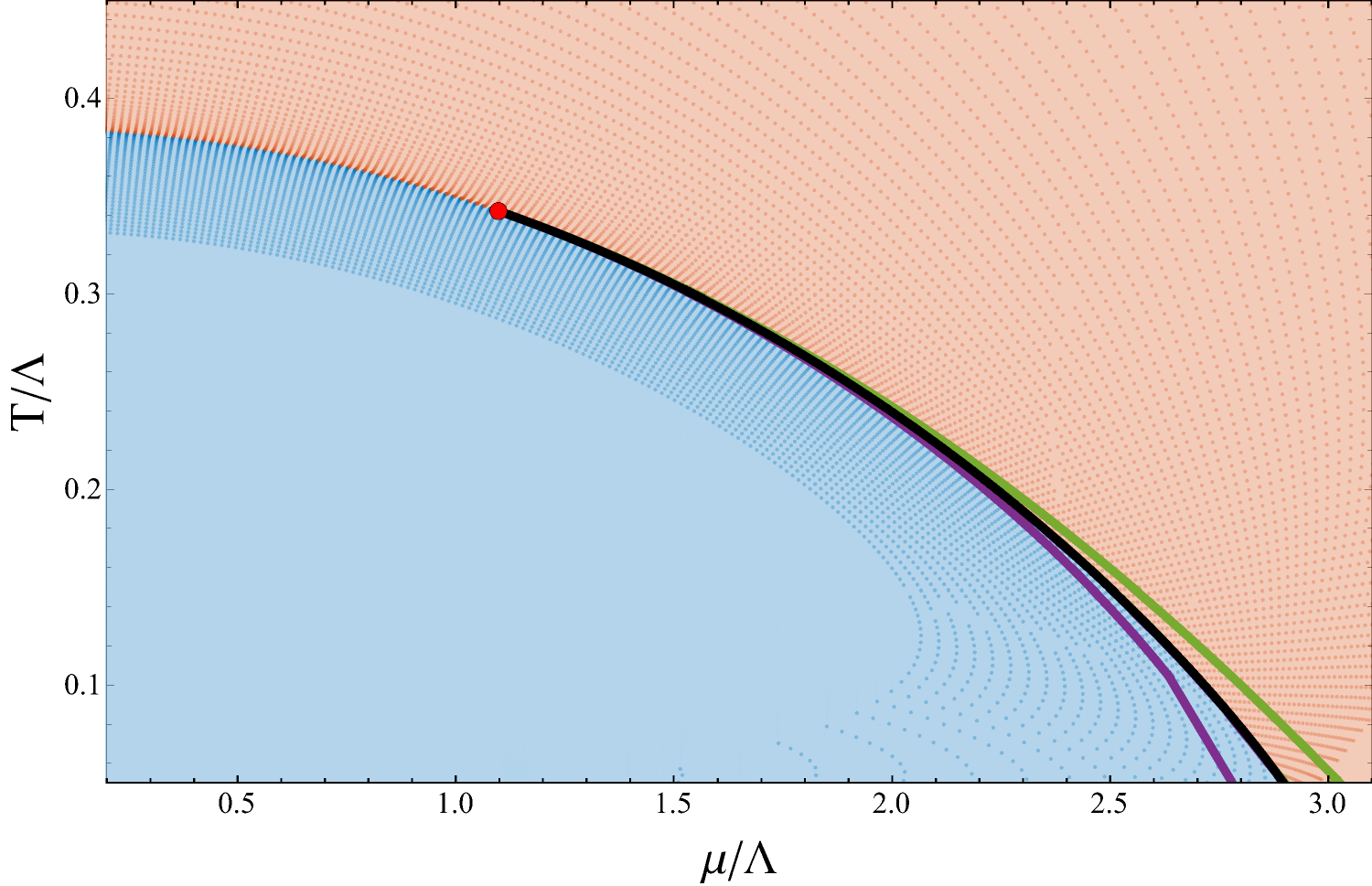}
\caption{Phase diagram. The black curve indicates the location of the FOPT. This curve terminates at the CP, marked by a red dot. The green and the purple curves indicate the boundaries of the metastable regions.}
\label{fig:phase_diagram}
\end{figure}

The phase diagram of the model is obtained by ``projecting'' e.g.~the energy density plot down onto the $(T,\mu)$ plane. The result is shown in \fig{fig:phase_diagram}.
Usually, only the projection of the globally stable states is included. However, in the figure we also show the boundaries of the metastable regions, since these  will play a crucial role in our analysis. The FOPT is indicated by a thick black line. As expected, the metastability regions in \fig{fig:phase_diagram} coincide with the multivalued regions in \fig{tmu_line_crossover}.

We will now explain how the stability properties are determined. First, we must identify the locally unstable states. For this purpose, we examine the susceptibility matrix, namely the Hessian of (minus) the free energy density
\be\label{eq:hessian}
    H=-\begin{pmatrix} 
    \dfrac{\partial^2 \mathcal{F}}{\partial T^2} & \dfrac{\partial^2 \mathcal{F}}{\partial\mu\partial T}\\[5mm]
    \dfrac{\partial^2 \mathcal{F}}{\partial T\partial \mu} & \dfrac{\partial^2 \mathcal{F}}{\partial \mu^2}\end{pmatrix} = \begin{pmatrix} \dfrac{\partial s}{\partial T} & \dfrac{\partial s}{\partial\mu}\\[5mm] \dfrac{\partial \rho}{\partial T} & \dfrac{\partial \rho}{\partial \mu}\end{pmatrix}\,.
\ee
In practice, we translate these derivatives to derivatives with respect to $\phi_H$ and $\Phi_H$, for which we constructed a regular grid. Locally unstable states are those for which the Hessian has at least one negative eigenvalue. The unstable region is the  yellow region in the figures above. 
As shown in Fig.~\ref{fig:slices_energy}, this region is related to the negative sign of the slope of the energy density when plotted against the temperature (chemical potential) at constant chemical potential (temperature). Precisely at the CP, the slope of the energy density  as a function of $T$ and $\mu$ diverges, as is characteristic of a second-order phase transition.
 
\begin{figure}[htbp]
\centering
    \includegraphics[width=.95\textwidth]{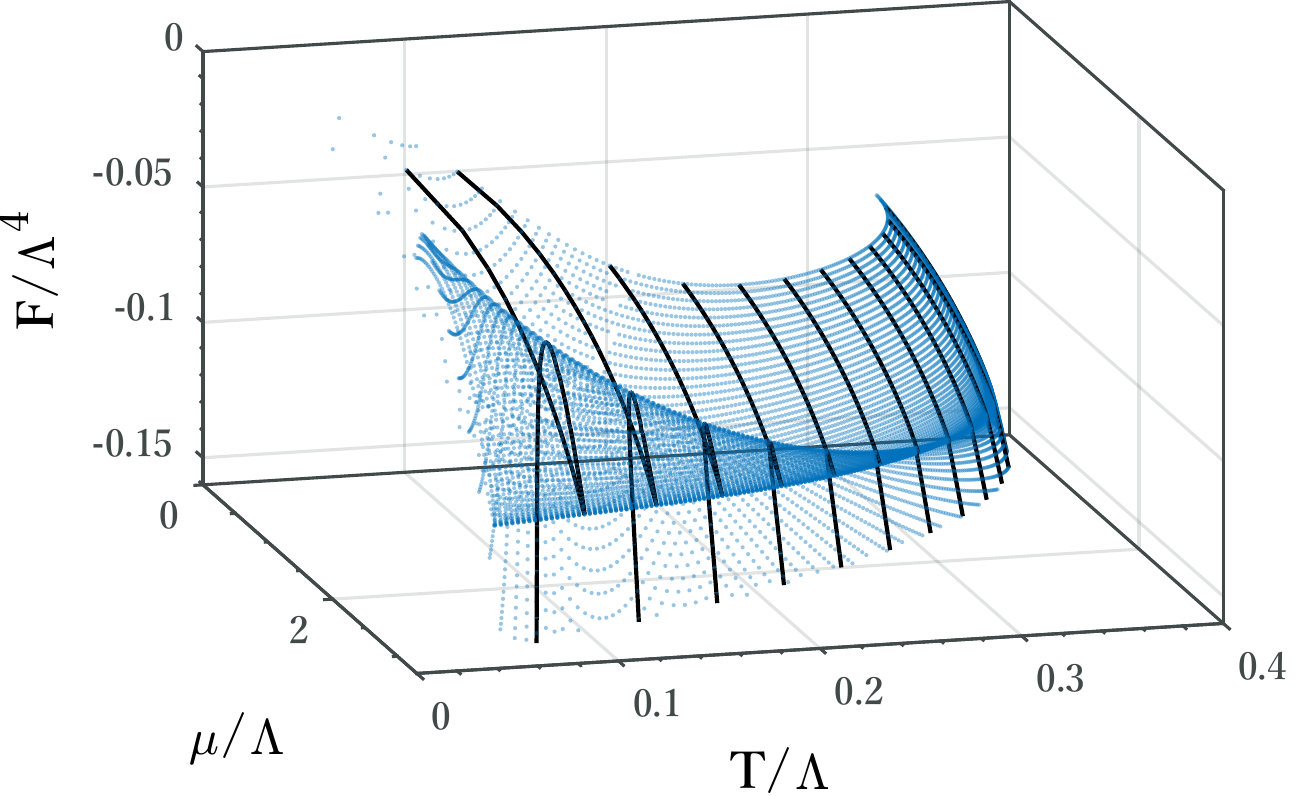}
\caption{Free energy density as a function of temperature and chemical potential. We show curves of constant $\Phi_H$  in black to make the swallow-tail structure clearer. \label{fig:swallow}}
\end{figure}

We now turn to metastable states. These are locally stable but not globally preferred. These states will play a crucial role regarding the dynamics of bubbles. To find them, we compute the free energy density as a function of $(T,\mu)$. The result is shown in \fig{fig:swallow}.
For each value of $(T,\mu)$, there are either one or three possible states. In the first case the state is directly the stable, globally preferred state. In the second case, the globally preferred state is that with the lowest value of the free energy. Out of the the other two states one is locally unstable and one is locally stable. This classification gives rise  the different colors in Figs.~\ref{fig:all_eq_solutions}, \ref{tmu_line_crossover} and \ref{fig:EvsTvsmu}. The  boundaries of the two globally stable regions (the orange and the blue regions) give rise to a single curve in the $(T,\mu)$ plane. This is the line of FOPT shown in black in \fig{fig:phase_diagram}.

\section{Evolution algorithm}
\label{time}

We are interested in studying the time evolution of expanding, planar bubbles along a chosen boundary direction $x$. Assuming translational and rotational invariance along the other two boundary directions, $y$ and $z$, the most general ansatz  in in-going 
Eddington-Finkelstein (EF) coordinates is \cite{Attems:2017zam} 
\begin{equation}
\begin{aligned}\label{eq:ansatz}
        ds^2 & = g_{\mu\nu}dx^{\mu}dx^{\nu}= -Adt^2+2dtdr+Fdtdx+S^2\Big[ e^{-2B}dx^2 + 
        e^{B}( dy^2+dz^2)\Big] \,,\\[2mm]
        A_{\mu}dx^{\mu} & = A_t dt + A_x dx \,.
\end{aligned}
\end{equation}
All the functions depend on $(t,r,x)$, with $r$ the holographic radial coordinate. Note that we are working in the gauge $A_r=0$.

The presence of the Maxwell field introduces minor modifications 
to the procedure followed in  \cite{Attems:2017zam, Jecco}. Details about these modifications, as well as about their  implementation in our numerical code \texttt{Jecco}, can be found in Appendix~\ref{sec:Jecco_Details}. Here we will just make  a few remarks. 

The fall-off of the different fields near the boundary, i.e.~as $r\rightarrow\infty$, takes the form
\begin{equation}
\begin{aligned}\label{eq:fall-off}
        & A = r^2 + \cdots + \frac{a_4(t,x)}{r^2}+\cdots ,\\[2mm]
        & B = \frac{b_4(t,x)}{r^4}+\cdots ,\\[2mm]
        & F = \cdots + f_2(t,x)u^2+\cdots ,\\[2mm]
        & \phi = \frac{\Lambda}{r}+ \cdots + \frac{\phi_2(t,x)}{r^3}+\cdots , \\[2mm]
        & A_t = A_{t,0} +  \frac{A_{t,2}(t,x)}{r}+\cdots ,\\[2mm]
        & A_x = A_{x,0} + \frac{A_{x,2}(t,x)}{r}+\cdots .
\end{aligned}
\end{equation}
We have only displayed the terms of direct interest here; the full expressions are given in Appendix~\ref{sec:Jecco_Details}. The leading-order terms in the gauge field series expansion,  $A_{t,0}$ and $A_{x,0}$, have the interpretation of a non-dynamical electromagnetic field at the boundary, which is absent in our case. Therefore we can and will work in  a gauge in which $A_{t,0}=A_{x,0}=0$. In other coordinate systems one  chooses a non-zero $A_{t,0}$ in order to ensure regularity at the bifurcation point of the black-brane horizon in the bulk. In those cases $A_{t,0}$ can be identified with the chemical potential of the boundary theory. In in-going EF coordinates the gauge field is regular at the future horizon with the choice $A_{t,0}=0$ and the chemical potential can be obtained as an integral along a curve connecting the boundary and the horizon --- see e.g.~\cite{Natsuume:2013lfa} for more details. 

The boundary expectation values of the stress tensor, the scalar operator and the baryon current are related to the fall-off coefficients through
\begin{equation}
    \begin{aligned}
        \mathcal{E} = \frac{\kappa_5^2}{2L^3} \langle T^{tt}\rangle & = -\frac{3}{4}a_4-\Lambda\phi_2+\left(\frac{7}{36}-\lambda_4\right)\Lambda^4,\\[2mm]
        \mathcal{P}_L = \frac{\kappa_5^2}{2L^3} \langle T^{xx}\rangle & = -\frac{a_4}{4}-2b_4+\frac{\Lambda\phi_2}{3}+\left(\frac{-5}{108}+\lambda_4\right)\Lambda^4,\\[2mm]
        \mathcal{P}_T = \frac{\kappa_5^2}{2L^3} \langle T^{aa}\rangle & = -\frac{a_4}{4}+b_4+\frac{\Lambda\phi_2}{3}+\left(\frac{-5}{108}+\lambda_4\right)\Lambda^4,\\[2mm]
        \mathcal{J} = - \frac{\kappa_5^2}{2L^3} \langle T^{tx}\rangle & = f_2,\\[2mm]
        \mathcal{V} = - \frac{\kappa_5^2}{2L^3} \langle \mathcal{O}_{\phi}\rangle & = -2\phi_2+\left(\frac{1}{3}-4\lambda_4\right)\Lambda^3,\\[2mm]
         \rho = \frac{\kappa_5^2}{2L^3} \langle j^t\rangle & = -2f(0)A_{t,2},\\[2mm]
         \mathcal{J}^Q = \frac{\kappa_5^2}{2L^3} \langle j^x\rangle & = 2f(0)A_{x,2},\\
    \end{aligned}\label{eq:VEVs}
\end{equation}
where the index ``$a$''  labels the transverse directions $y$ and $z$. The normalization factor is related to the number of degrees of freedom at the boundary: $2L^3/\kappa_5^2 \propto N^2$. The stress tensor and the baryon current obey the conservation laws
\be
\partial_{\mu}\langle T^{\mu\nu} \rangle = 0,\quad \quad \partial_{\mu}\langle j^{\mu} \rangle = 0,
\label{eq:conservation_laws}
\ee
while the trace of the stress-tensor is related to the scalar expectation value through the Ward identity
\be
\langle T^{\mu}_{\mu}\rangle = -\Lambda\langle \mathcal{O}_{\phi}\rangle.
\ee


In order to evolve the system in time, we start by specifying the bulk  initial  data $\phi(0,r,x)$, $B(0,r,x)$ and $A_x(0,r,x)$, together with the boundary initial data $a_4(0,x)$, $f_2(0,x)$ and $A_{t,2}(0,x)$. We then apply the algorithm described in Appendix~\ref{sec:Jecco_Details} to evolve to the next time step. At each time slice we read off the coefficients \eqref{eq:fall-off} and use \eqref{eq:VEVs} to compute the expectation values of the boundary operators.

\section{Bubble dynamics}
Our analysis here extends that of \cite{Bea:2021zsu}, which we follow closely. 

We construct initial conditions as follows. 
We consider a homogeneous, metastable  state with a large, localized perturbation around the origin. We implement this in \texttt{Jecco} \cite{Bea:2022mfb} by choosing a Gaussian perturbation in the $a_4$ and $A_{t,2}$  fall-off coefficients. The two possible signs of the perturbation correspond to initial conditions characterized by either a localized excess or deficit of energy and charge. 
To simulate superheated bubbles, we choose the metastable state to be in the green region of Figs.~\ref{fig:all_eq_solutions} and \ref{fig:EvsTvsmu}, and the perturbation to correspond to an excess of energy and charge  around $x=0$. To simulate supercooled bubbles, we  choose the metastable state to be in the purple region of Figs.~\ref{fig:all_eq_solutions} and \ref{fig:EvsTvsmu}, and the perturbation to correspond to a deficit of energy and charge  around $x=0$. If the amplitude of the perturbation is large enough, the subsequent dynamics turns the perturbation into an expanding bubble that quickly reaches a terminal velocity. 

\begin{figure}[tbp]
\centering
    \includegraphics[width=.95\textwidth]
    {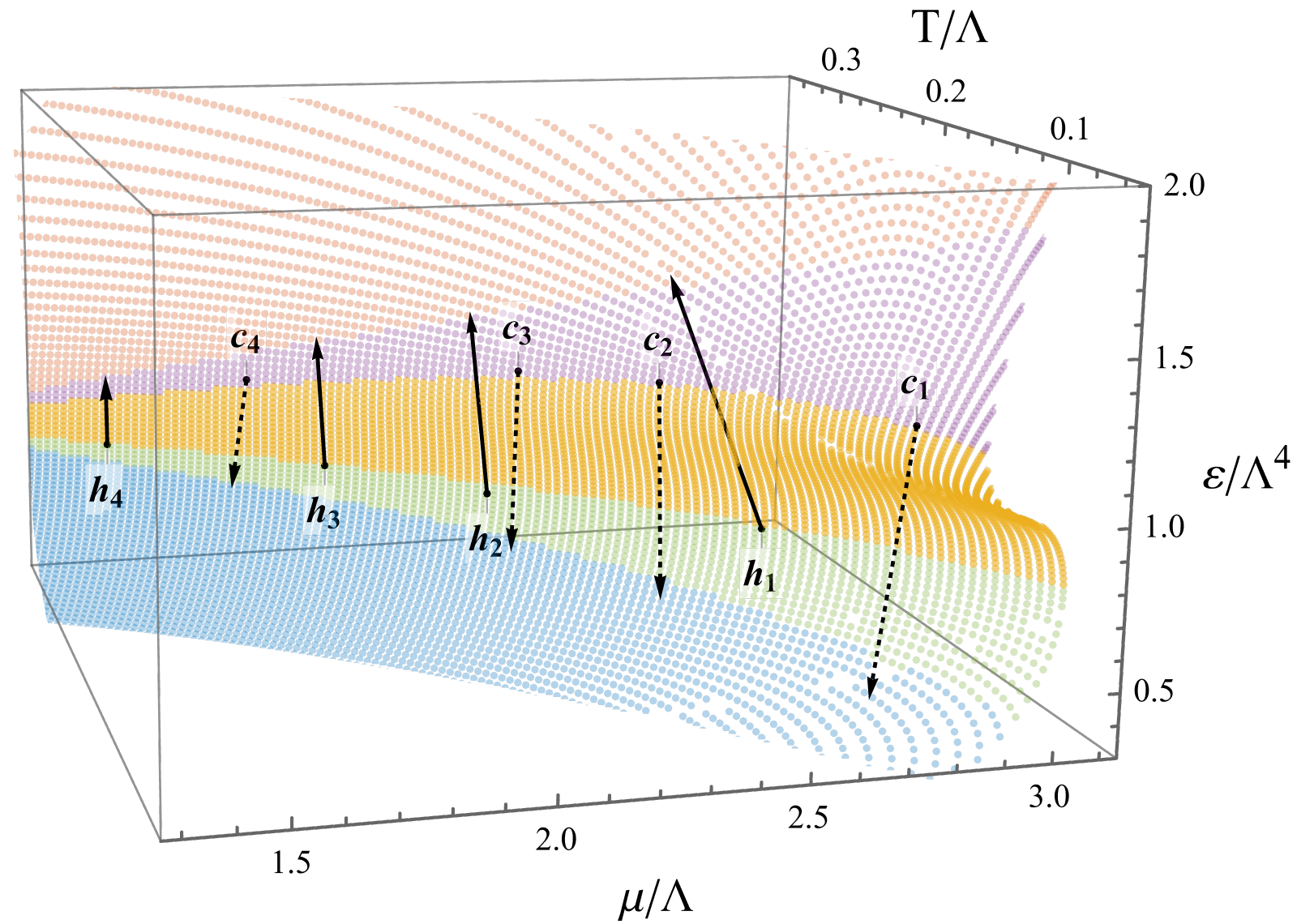}\\[10mm]
    \includegraphics[width=.95\textwidth]
    {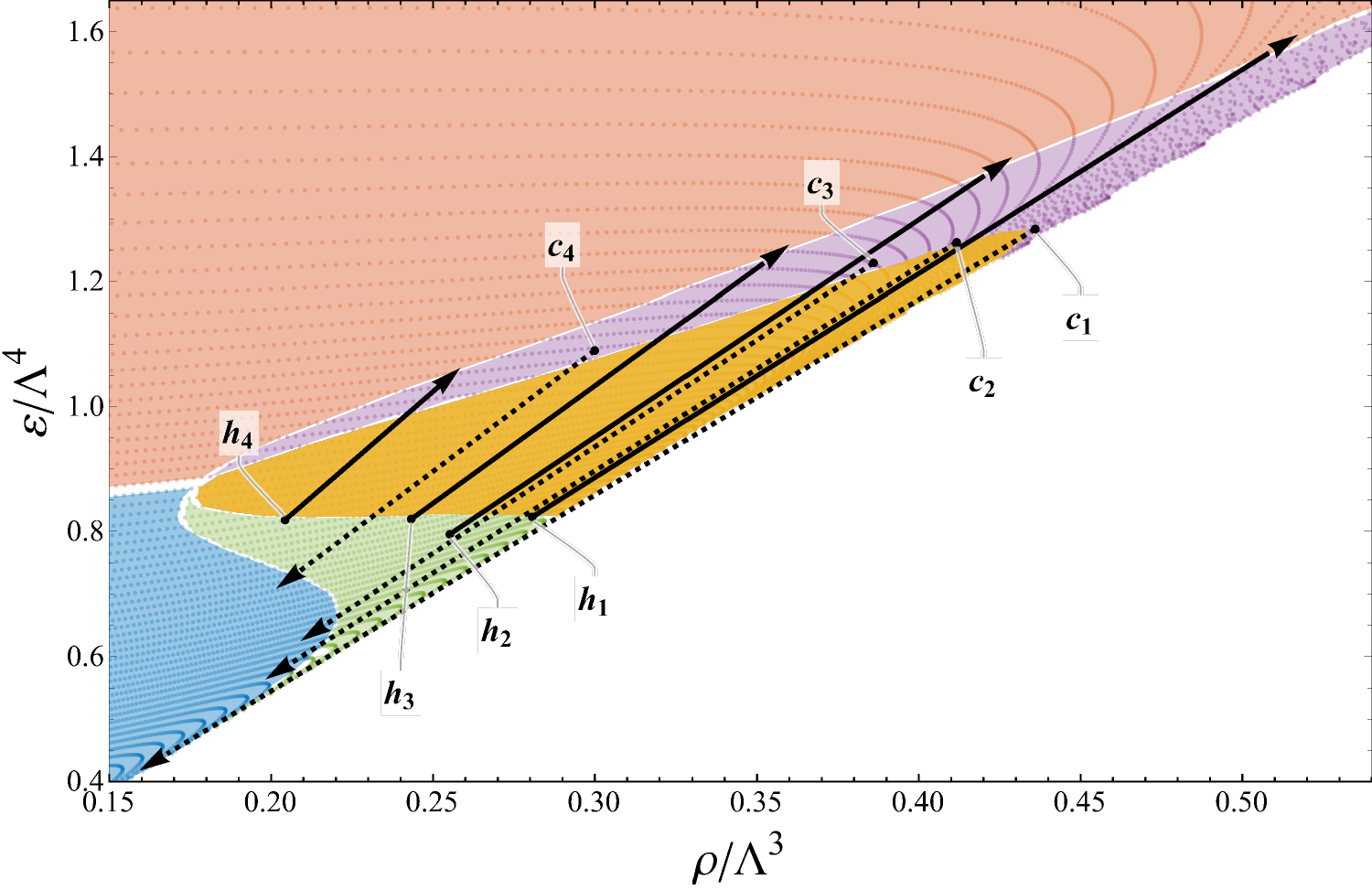}
\caption{Each arrow in these plots represents the properties of an expanding bubble once it has reached a steady state. The tail of the arrow  indicates the values of $(T,\mu,\mathcal{E})$ (top) or $(\mathcal{E}, \rho)$ (bottom) in the metastable state outside the bubble. The head of the arrow corresponds to the values in the stable state inside the bubble. As is clear from both plots, solid and dashed arrows correspond to superheated ($h_i$) and supercooled ($c_i$) bubbles, respectively. The arrows shown represent a few representative bubbles from the larger subset that we studied.}
\label{fig:inner_state}
\end{figure}

We conducted multiple time-evolution simulations starting from various initial metastable states, resulting in a series of superheated and supercooled bubbles. These simulations were executed using the Booster partition of the LEONARDO \cite{leonardo} supercomputer to parallelize the computations. Each of these simulations can be represented by an arrow in \fig{fig:inner_state}, where we only show a few representative examples. The tail of the arrow indicates the  metastable state outside the bubble. This can be freely specified  and  characterizes the bubble in question. The head of the arrow indicates the stable state inside the bubble. This cannot be freely specified: once the metastable state is chosen, both the state inside the bubble and the bubble wall velocity are determined dynamically  \cite{Bea:2021zsu}. In particular, although the interior of the  bubble at nucleation is typically assumed to have the same values of $(T,\mu)$ as the metastable state,

these values differ once the steady state is achieved.

\begin{figure}[tbp]
\centering
{\includegraphics[width=.8\textwidth]{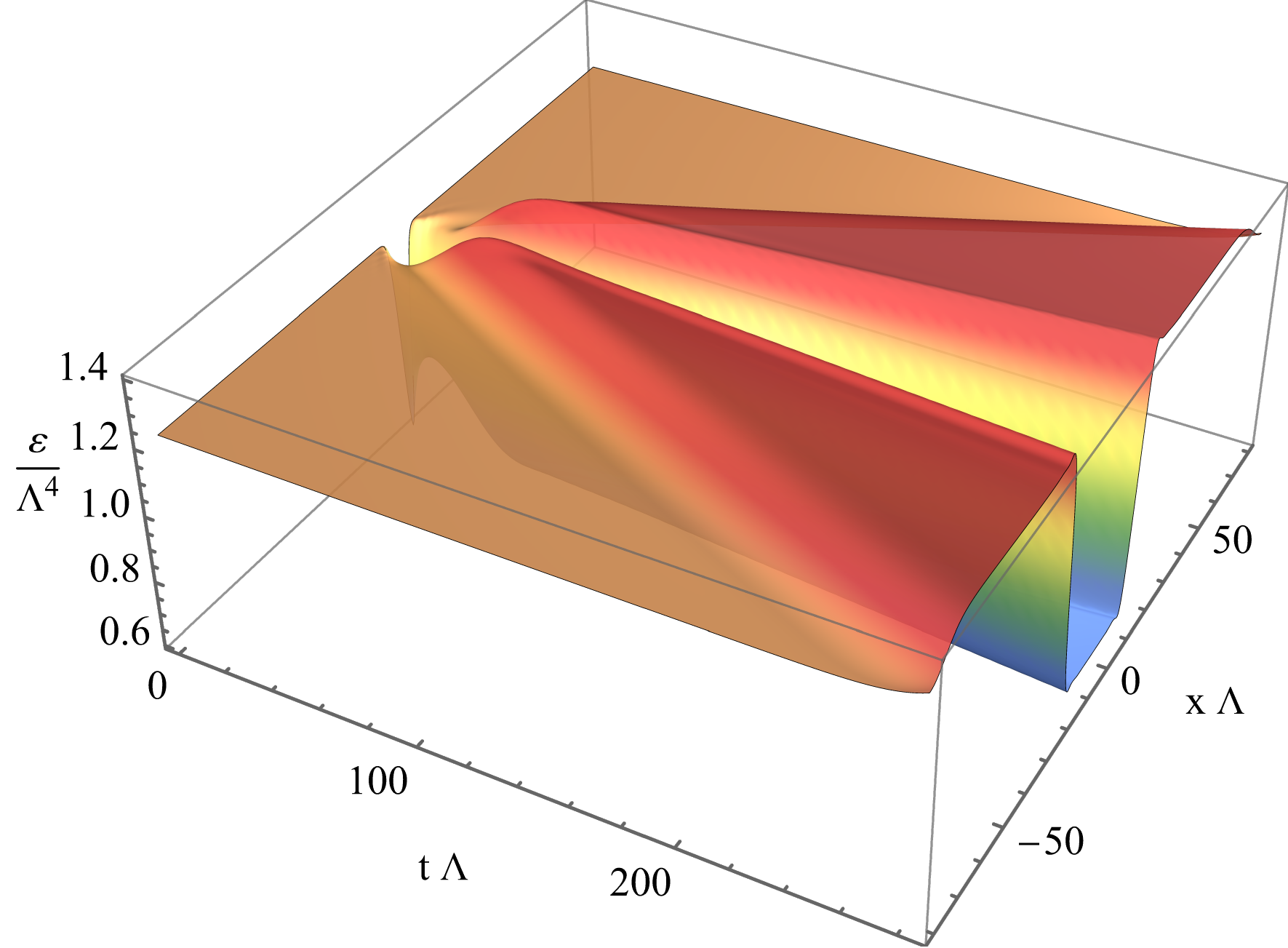}\\[10mm]
\includegraphics[width=.8\textwidth]{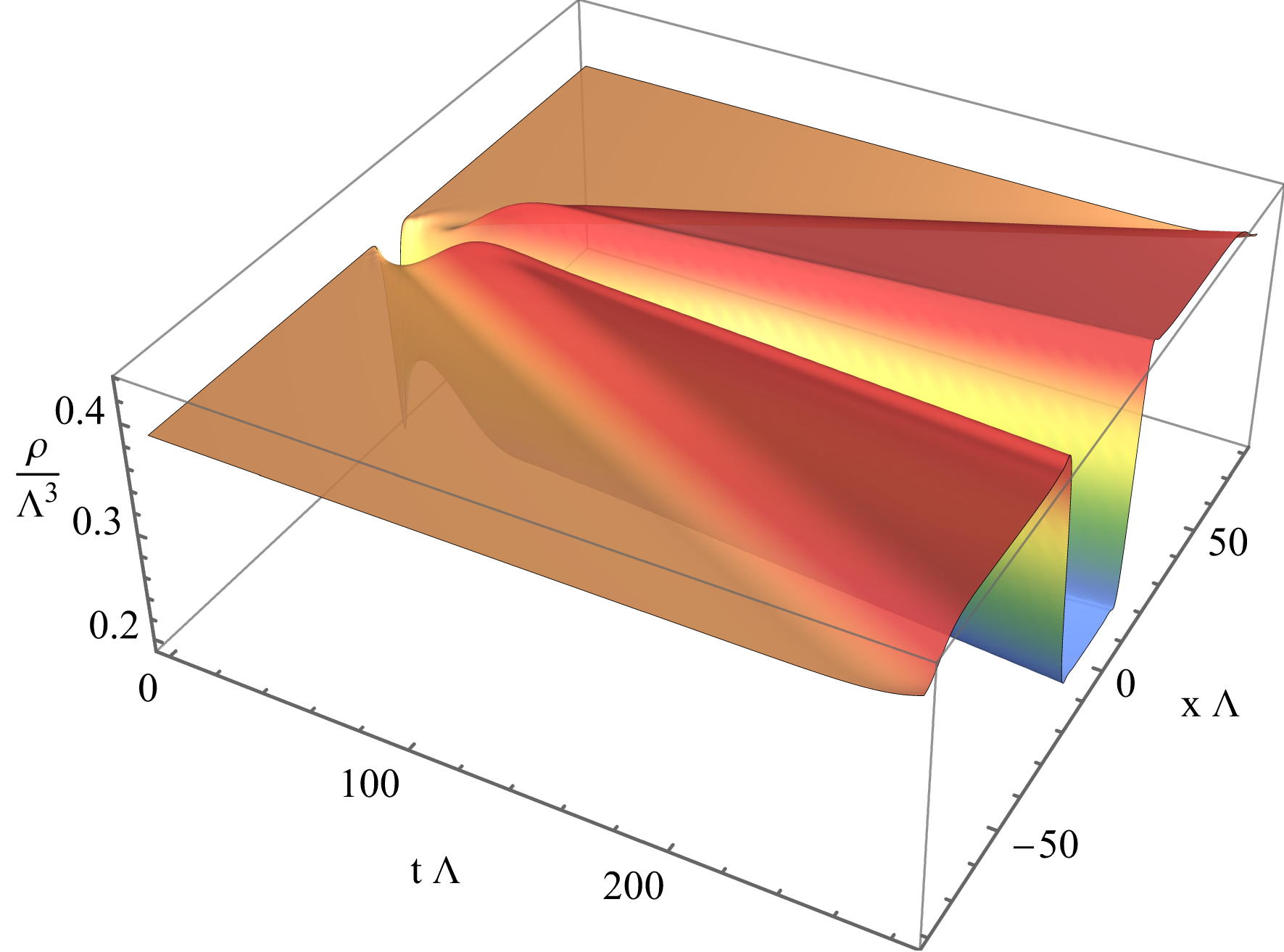}}
\caption{Time evolution the energy (top) and the baryon (bottom) densities for an expanding supercooled bubble corresponding to label $c_3$ of \fig{fig:inner_state}.}
\label{fig:supercooled_3D}
\end{figure}

\begin{figure}[tbp]
\centering
{\includegraphics[width=.8\textwidth]{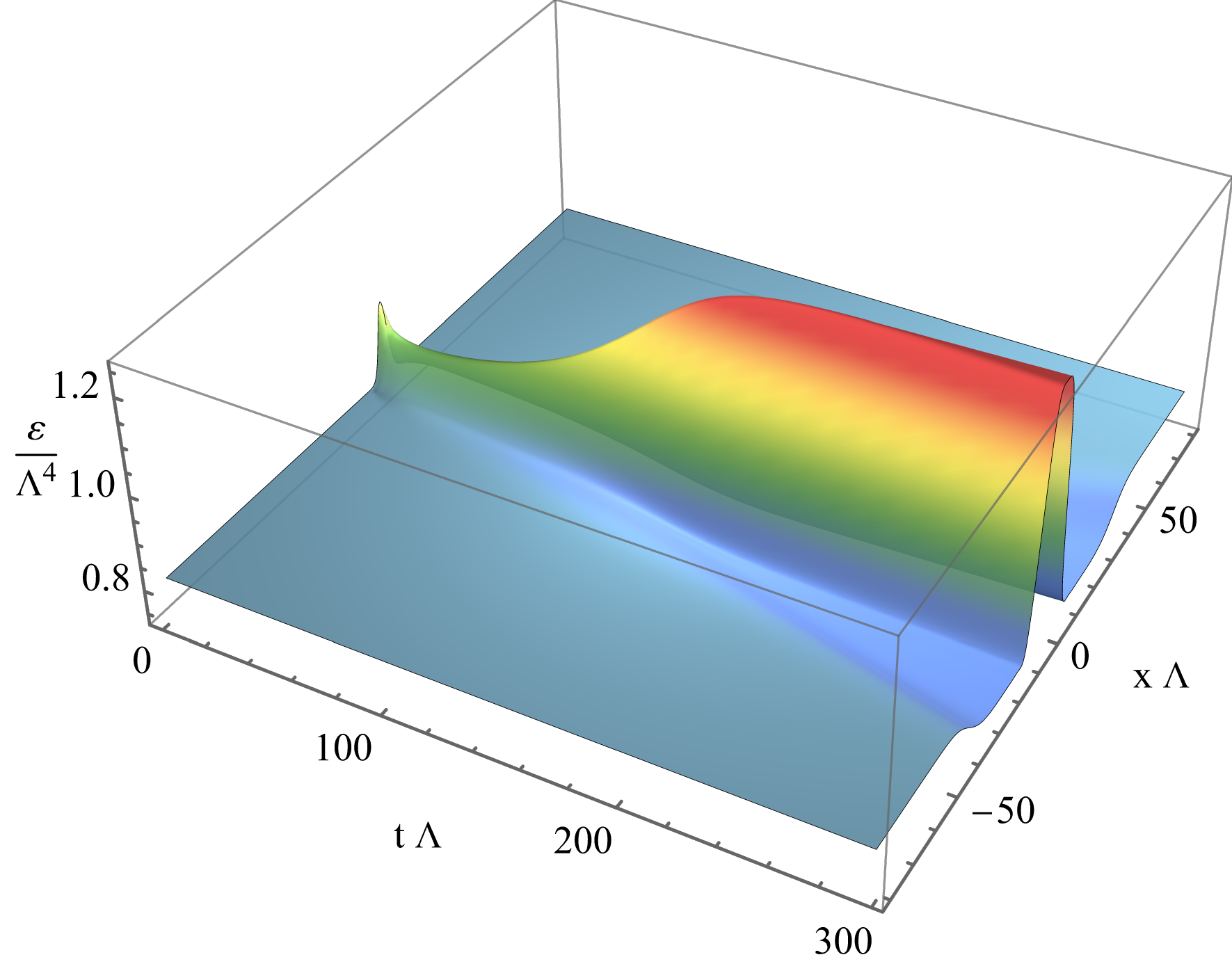}\\[10mm]
\includegraphics[width=.8\textwidth]{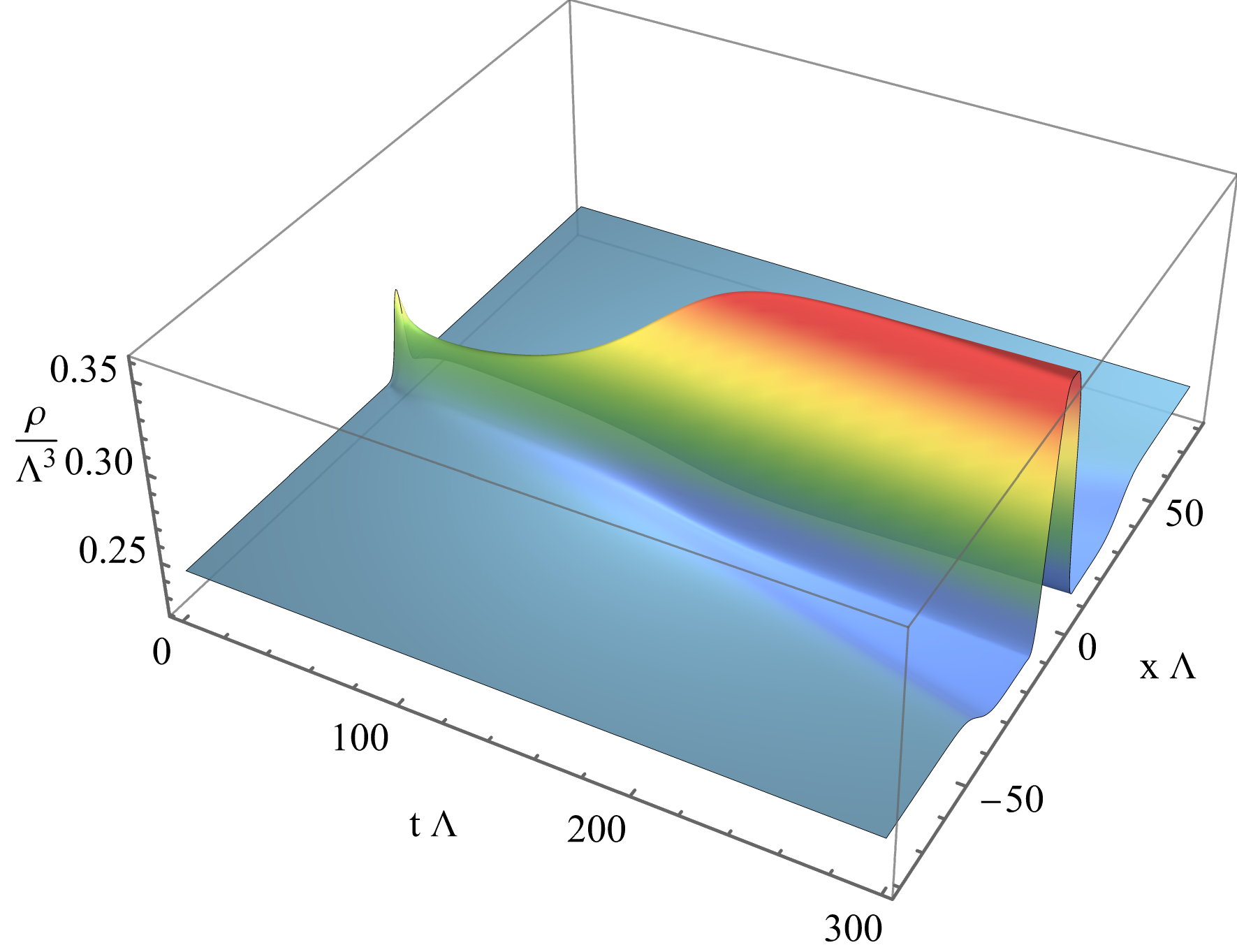}}
\caption{Time evolution the energy (top) and the baryon (bottom) densities for an expanding superheated bubble corresponding to label $h_3$ of \fig{fig:inner_state}.}
\label{fig:superheated_3D}
\end{figure}

\begin{figure}[tbp]
\centering
{\includegraphics[width=.47\textwidth]{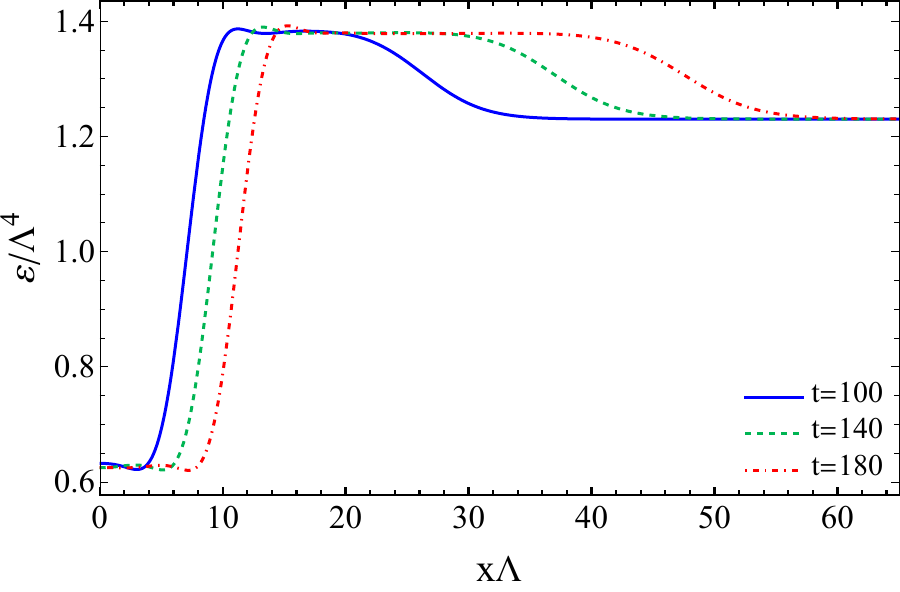} \hspace{4mm}
\includegraphics[width=.47\textwidth]{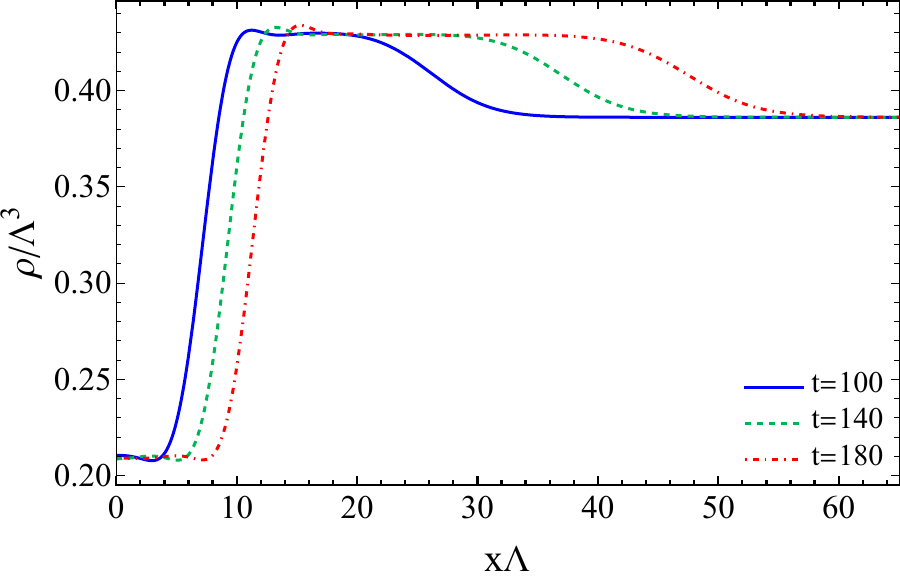}}\\[4mm]
{\includegraphics[width=.47\textwidth]{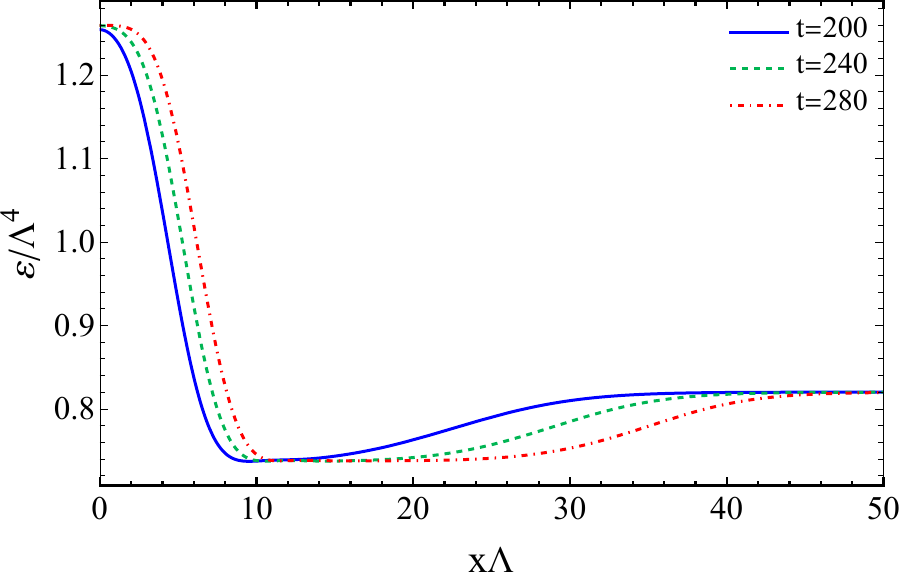}\hspace{4mm}
\includegraphics[width=.47\textwidth]{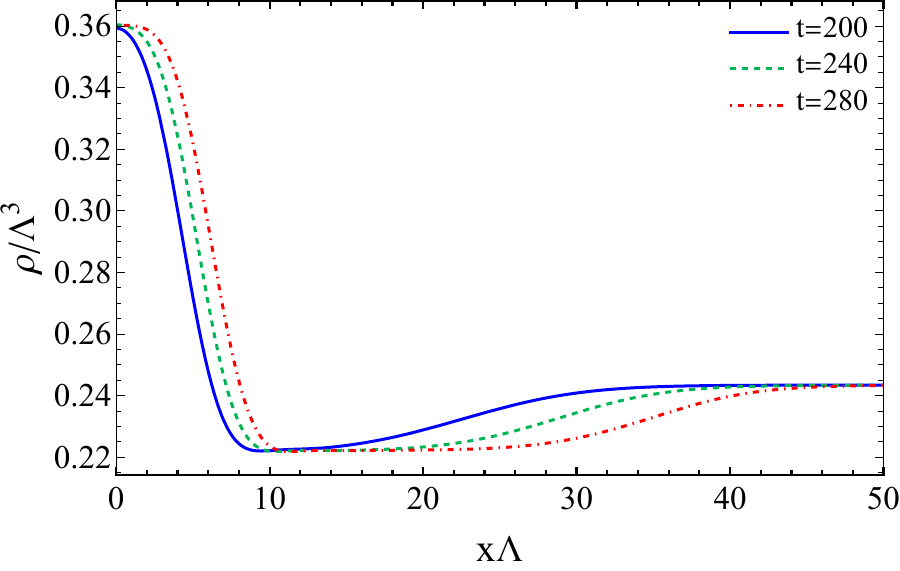}}
\caption{Snapshot of the energy (left) and baryon (right) densities for the supercooled (top) and superheated (bottom) bubbles of Figs.~\ref{fig:supercooled_3D} and~\ref{fig:superheated_3D}.}
\label{snapshots}
\end{figure}

Representative examples of the time evolution of a supercooled and a superheated bubble are shown in Figs.~\ref{fig:supercooled_3D} and~\ref{fig:superheated_3D}, respectively. Snapshots of the energy density and the baryon density are shown in \fig{snapshots}. Both cases corresponds to deflagrations. 
As noted in the study of their hydrodynamic properties \cite{Barni:2024lkj,Bea:2024bxu}, there is a crucial qualitative difference between supercooled and superheated bubbles. Supercooled bubbles push the fluid outward and build an overdense shell in front of them. In contrast, superheated bubbles absorb energy from the outside metastable state and build an underdense shell in front of them. These features are clearly visible in Figs.~\ref{fig:supercooled_3D}, \ref{fig:superheated_3D} and \ref{snapshots}.

\begin{figure}[tbp]
\centering
{\includegraphics[width=.80\textwidth]{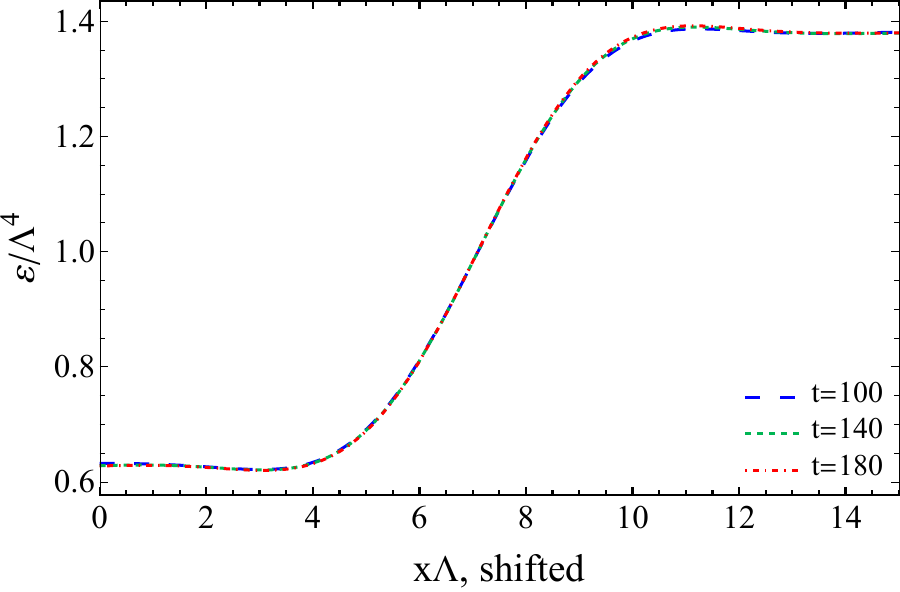}\\[10mm]
\includegraphics[width=.80\textwidth]{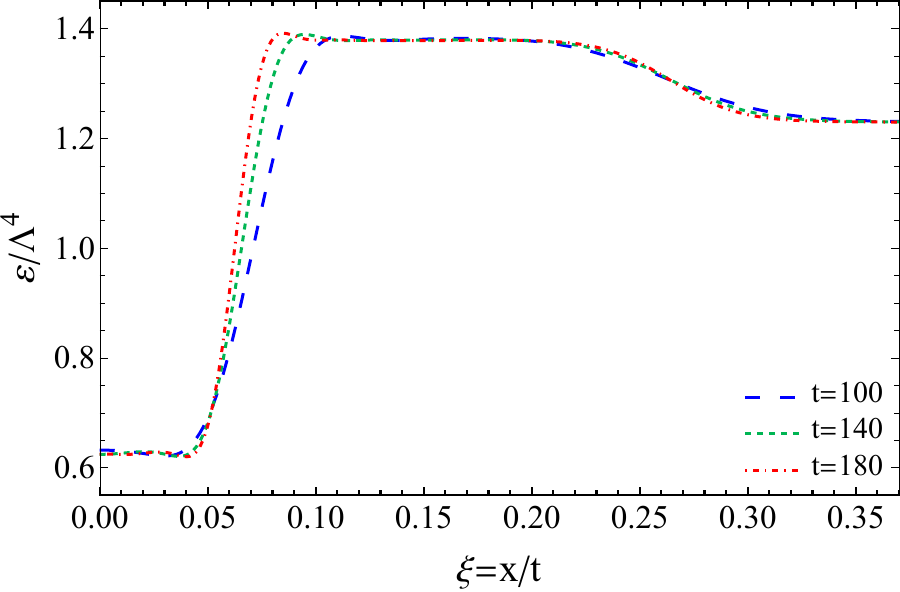}}
    \caption{(Top) Snapshots of the energy density for the supercooled bubble of \fig{fig:supercooled_3D} at different times. To illustrate  the agreement, different  snapshots have been shifted in $x$ by different constant amounts. (Bottom) Snapshots of the energy density for the same bubble at different times plotted as a function of the self-similar parameter, $\xi = x/t$.}
\label{fig:selfsim_hot}
\end{figure}

\begin{figure}[tbp]
\centering
{\includegraphics[width=.80\textwidth]{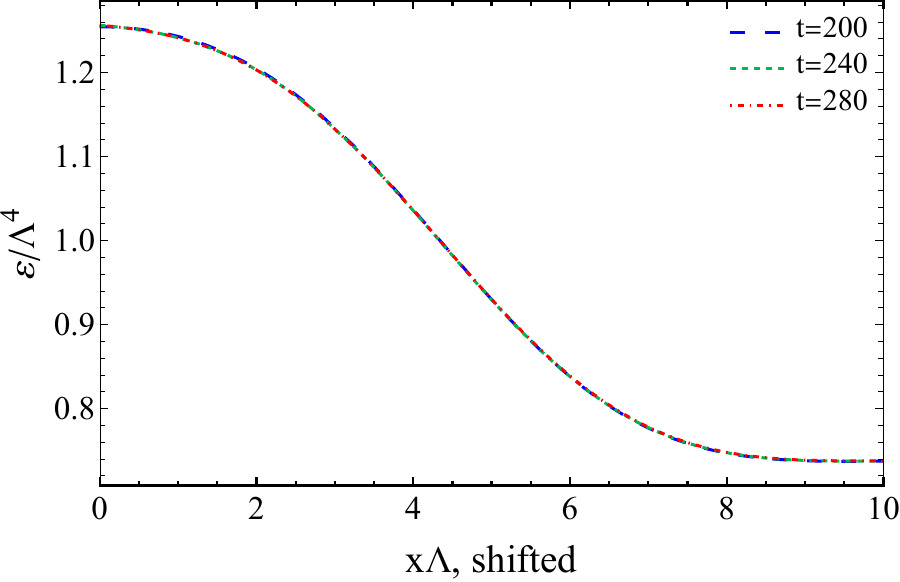}\\[10mm]
\includegraphics[width=.80\textwidth]{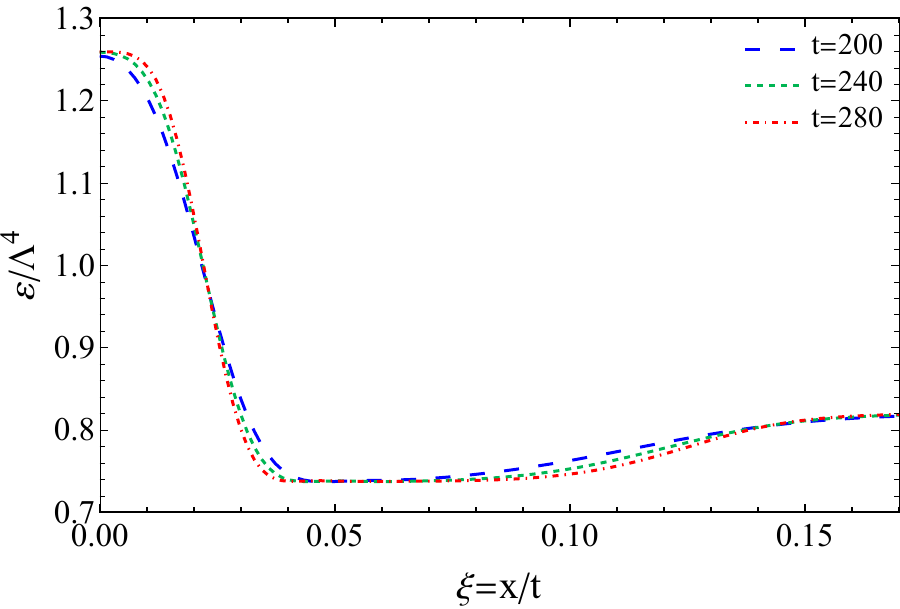}}
    \caption{(Top) Snapshots of the energy density for the superheated bubble of \fig{fig:superheated_3D} at different times. To illustrate  the agreement, different  snapshots have been shifted in $x$ by different constant amounts. (Bottom) Snapshots of the energy density for the same bubble at different times plotted as as a function of the self-similar parameter, $\xi = x/t$.
    }
\label{fig:selfsim_cold}
\end{figure}

As the steady-state regime is approached, the bubble wall profile becomes time-independent, as shown in \fig{fig:selfsim_hot}(top) and \fig{fig:selfsim_cold}(top). In this limit, the fluid flow approaches a self-similar profile. This is illustrated in \fig{fig:selfsim_hot}(bottom) and \fig{fig:selfsim_cold}(bottom), where different snapshots of the energy density as a function of the self-similar parameter seem to converge to a limiting profile.
Ref.~\cite{Bea:2021zsu} showed that the steady state reached by the bubble is completely independent of the initial conditions except for the metastable state outside the bubble. We expect the same conclusion to hold here. While we have not performed an analysis as exhaustive as that in Ref.~\cite{Bea:2021zsu}, our initial investigations agree with this expectation.  

\section{Wall velocity}
\begin{figure}[tbp]
\centering
\raisebox{-0.5\height}{\includegraphics[width=.82\textwidth]{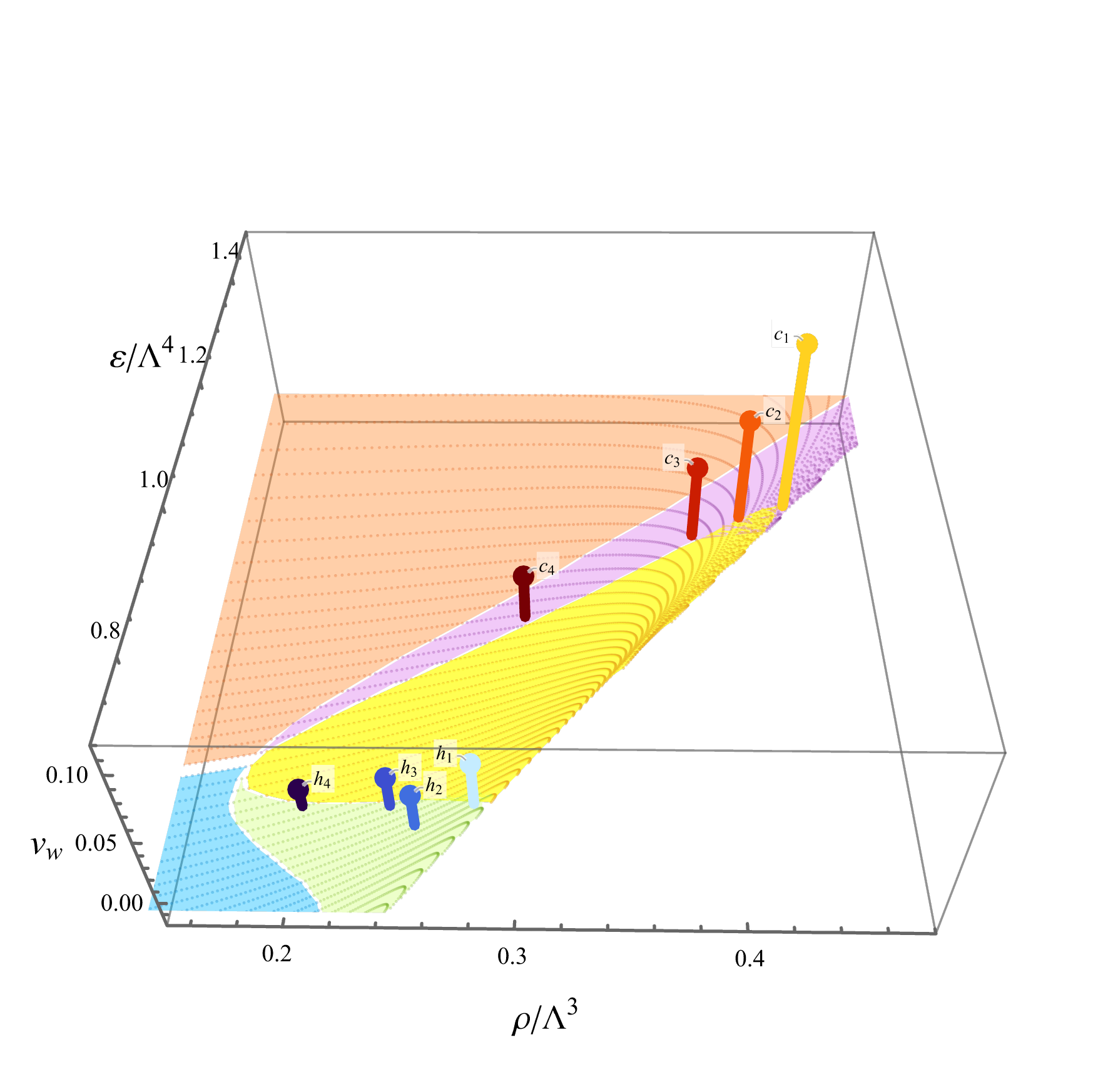}} \hspace{2mm}
\raisebox{-0.5\height}{\includegraphics[height=35mm]{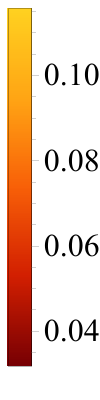}}  \hspace{1.5mm}
\raisebox{-0.5\height}{\includegraphics[height=35mm]{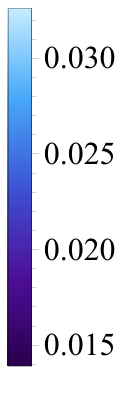}}
\caption{Bubble wall velocity for bubbles nucleated in different metastable states, labeled as in \fig{fig:inner_state}. The red and blue colorbars correspond to supercooled and superheated bubbles, respectively.}
\label{energychargevelocity}
\end{figure}

As illustrated in the top panels of Figs.~\ref{fig:selfsim_hot} and \ref{fig:selfsim_cold}, in the steady-state regime the wall moves as a rigid object. We define its position as the position of the  inflection point in the energy density profile, see \fig{snapshots}; other definitions lead to similar results. Monitoring the location of this point as a function of time we extract the bubble wall velocity, $v_w$, for each simulation. Repeating this process for many different simulations we obtain the dependence of the wall velocity on the metastable state on which it is nucleated, namely, on the values of $(T,\mu)$ or, equivalently,  $(\mathcal{E},\rho)$. In \fig{energychargevelocity}, we present the results for the same transitions considered in \fig{fig:inner_state}.
Within our set of simulations, the maximum velocity that we obtained was 
\begin{align}
\label{vmax}
    v_w^{\textrm{max}} \simeq 0.126 \quad \text{(supercooled)}\,, \\
    v_w^{\textrm{max}} \simeq 0.032 \quad \text{(superheated)}\,.
\end{align}

\begin{figure}[htbp]
\centering
\includegraphics[width=.95\textwidth]{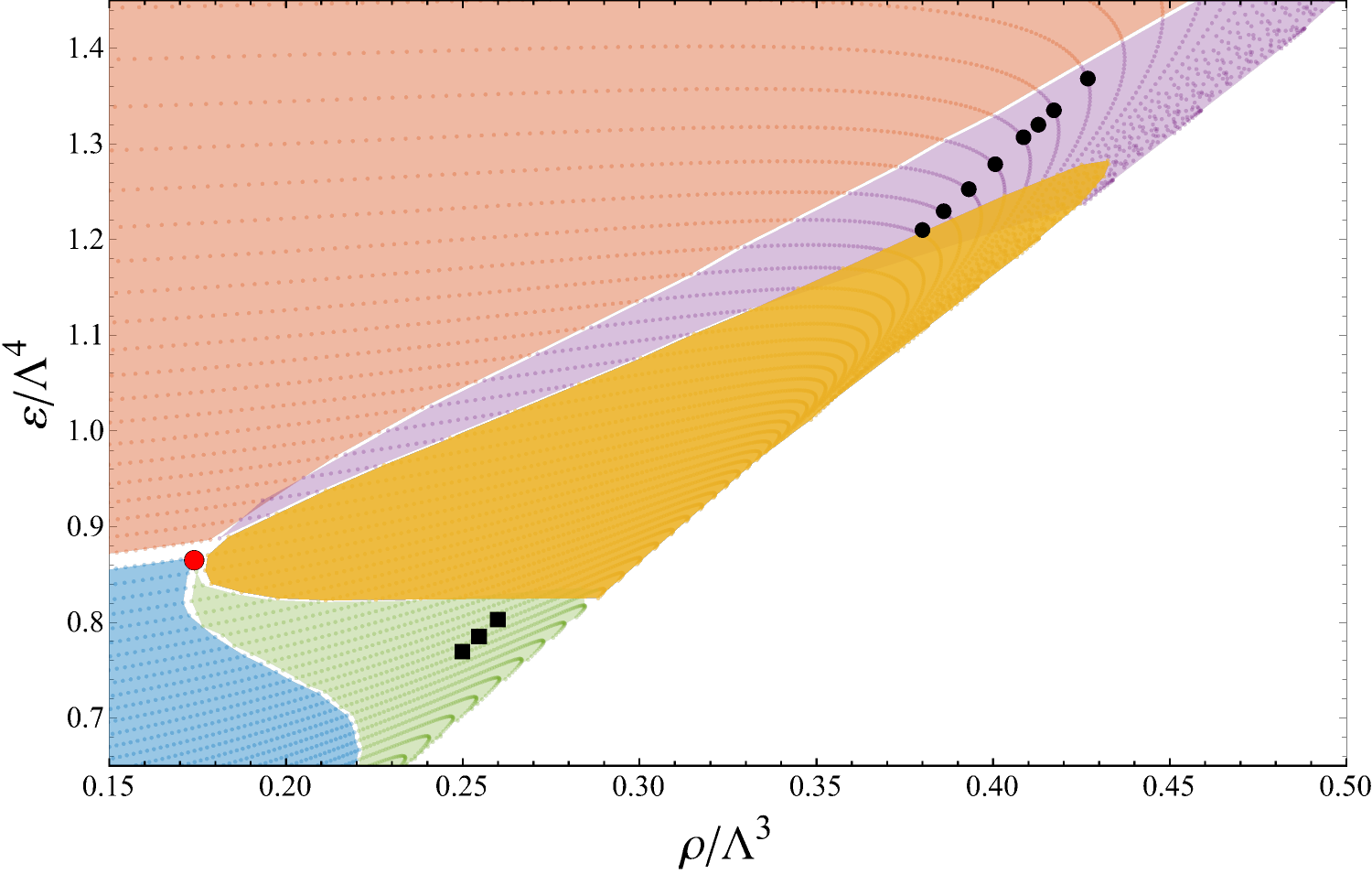}\\[10mm]
\includegraphics[width=.95\textwidth]{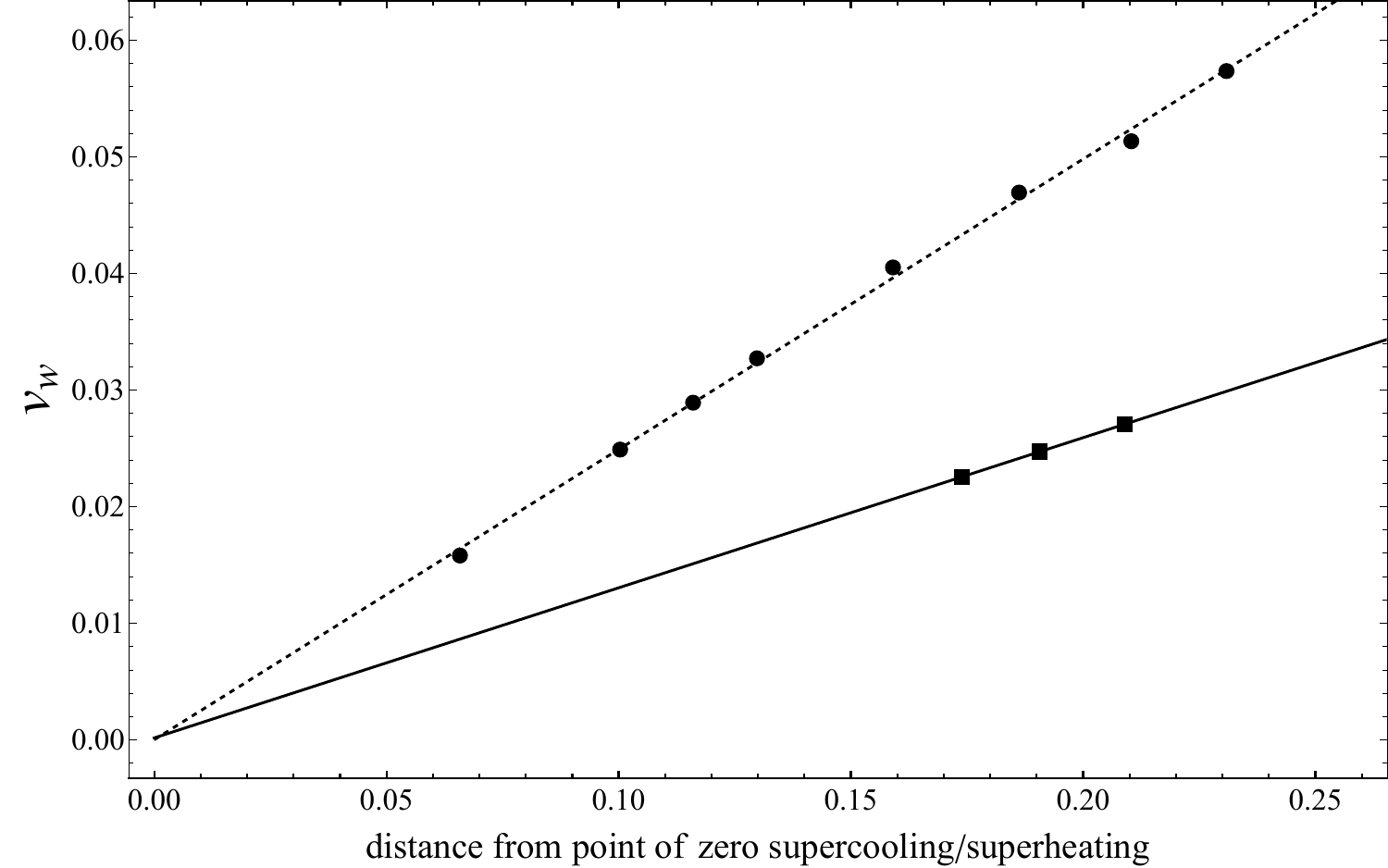}
\caption{(Top) Points in metastable regions on the $(\mathcal{E},\rho)$-plane. The black circles and squares  correspond to supercooled and superheated bubbles, respectively. The red circle marks the CP. (Bottom) Bubble wall velocity as a function of the distance to the point of zero supercooling or superheating. The solid and dashed lines correspond to linear fits to the superheated and supercooled data, respectively.}
\label{fig:vw_vs_distance_to_turning}
\end{figure}

The wall velocity increases monotonically with the amount of supercooling or superheating. To illustrate this feature, in  
\fig{fig:vw_vs_distance_to_turning}(top) 
we consider two sets of points along two (approximate) straight lines in the $(\mathcal{E},\rho)$-plane. 
The circles  correspond to supercooled bubbles. They start at a point of maximal supercooling (on the boundary between unstable and metastable states) and end close to a point of zero supercooling (on the boundary between stable and metastable states). The squares  correspond to superheated bubbles with different degrees of superheating.
The points of maximal supercooling and superheating  correspond to  turning points in \fig{fig:slices_energy}. The points of zero supercooling or superheating are points on the line of FOPT, namely, on the black curve in the phase diagram \fig{fig:phase_diagram}. The velocities corresponding to these two sets of points are plotted in \fig{fig:vw_vs_distance_to_turning}(bottom) as a function of the distance to the turning points, defined as the Euclidean distance in the coordinates $(\mathcal{E}/\Lambda^4, \rho/\Lambda^3)$. This distance does not have an intrinsic physical meaning but it provides a convenient operational manner to order the points. As expected, the velocity decreases monotonically and goes to zero at the point where the supercooling or the superheating vanishes. It would be interesting to determine whether the approximate linear behavior observed in \fig{fig:vw_vs_distance_to_turning}(bottom) is due to a physical effect or merely a consequence of the limited range of velocities considered.

\begin{figure}[htbp]
\centering
\includegraphics[width=.95\textwidth]{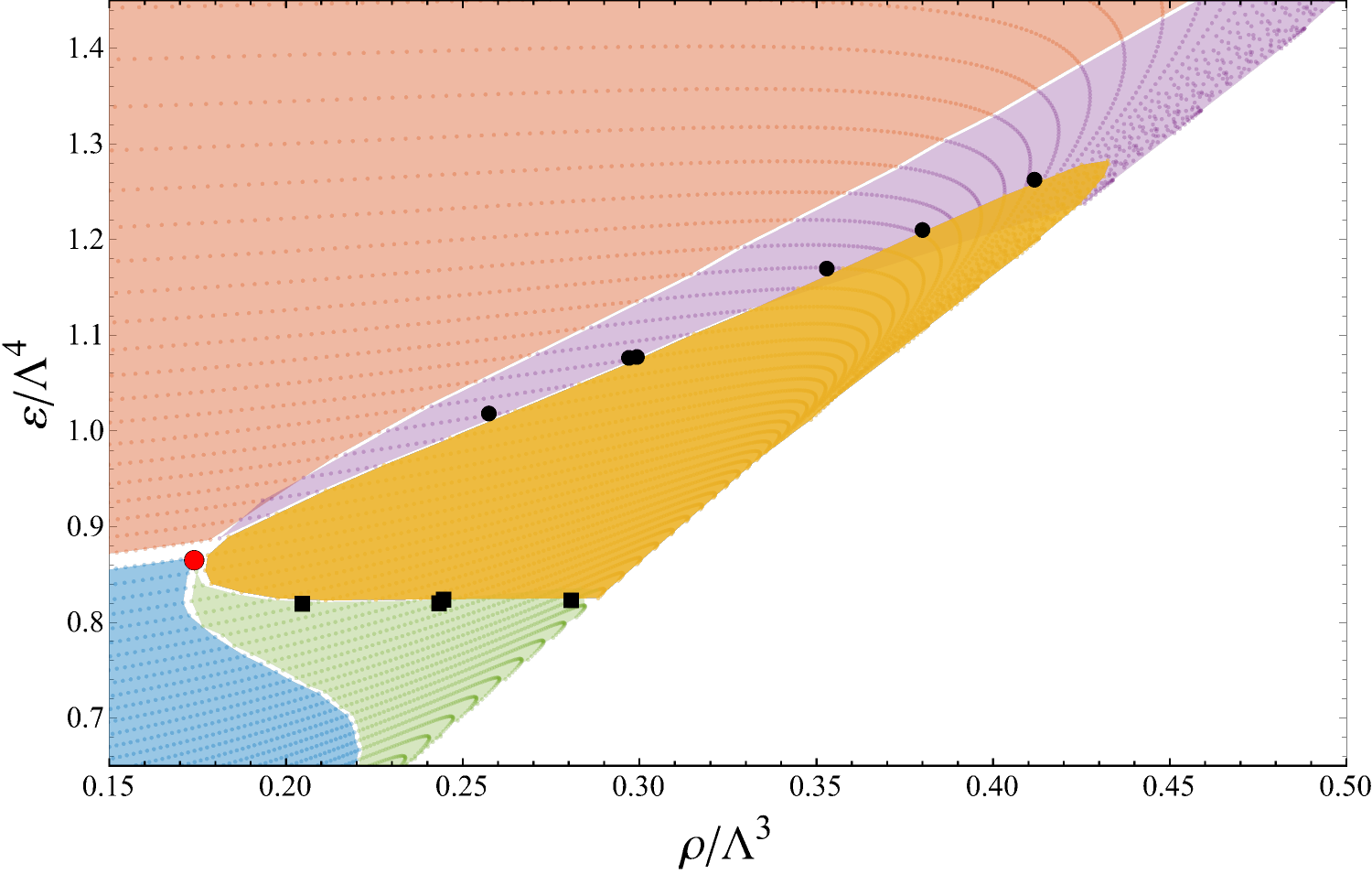}\\[10mm]
\includegraphics[width=.95\textwidth]{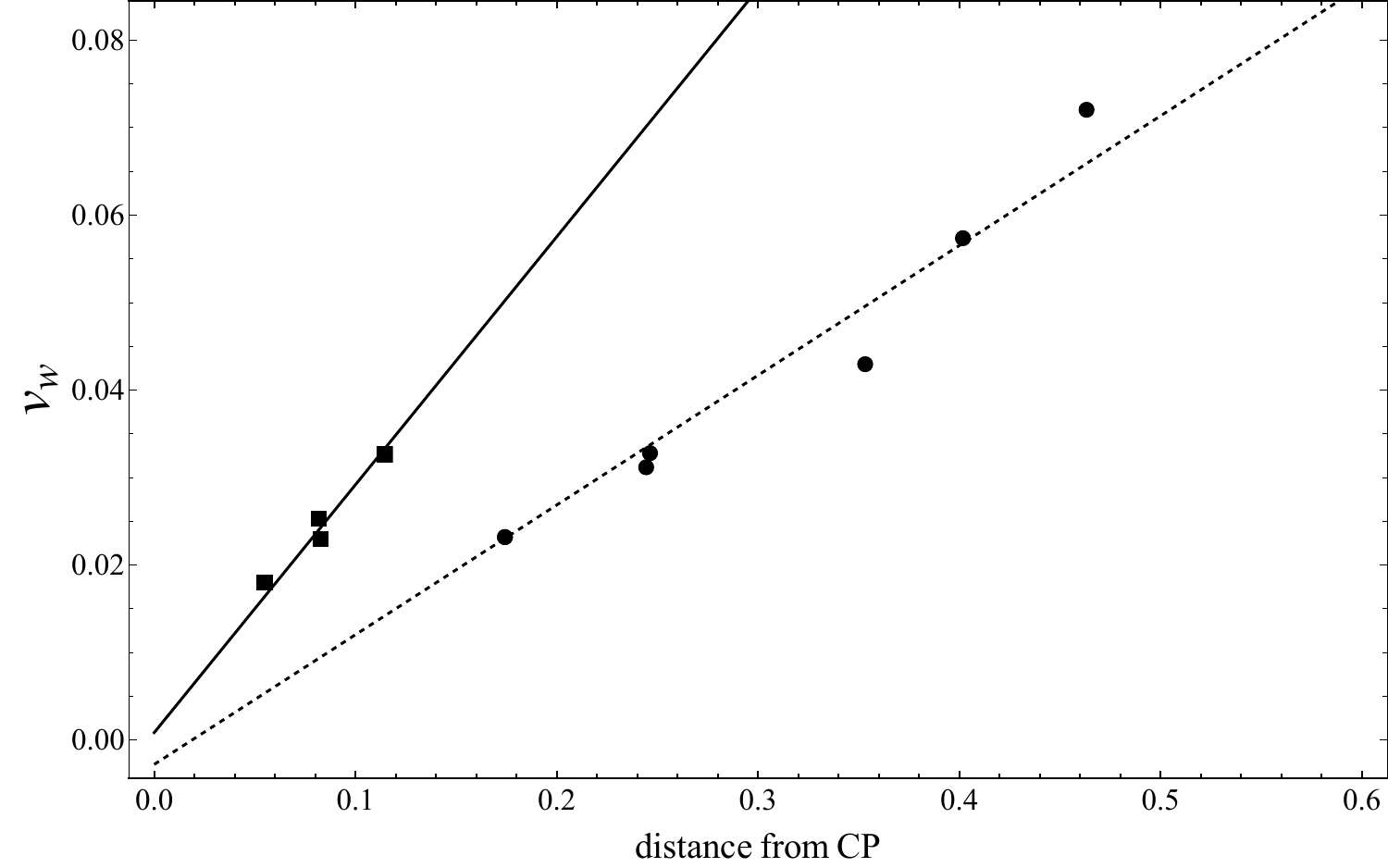}
\caption{(Top) Points in the $(\mathcal{E},\rho)$-plane on the curves of maximal supercooling (black circles) and maximal superheating (squares). The red circle marks the CP. (Bottom) Wall velocity as a function of the distance to the CP. The solid and dashed lines correspond to linear fits to the superheated and supercooled data, respectively.}
\label{fig:vw_vs_distance_to_CP}
\end{figure}

Consider now several points on the boundary of the metastable and unstable regions, namely, points of maximal supercooling or superheating, as those in \fig{fig:vw_vs_distance_to_CP}(top). In this case we expect that the wall velocity increases monotonically with the distance to the CP, since at the CP the FOPT disappears. This expectation is confirmed by \fig{fig:vw_vs_distance_to_CP}(bottom). 

\begin{figure}[tbp]
\centering
\includegraphics[width=.48\textwidth]{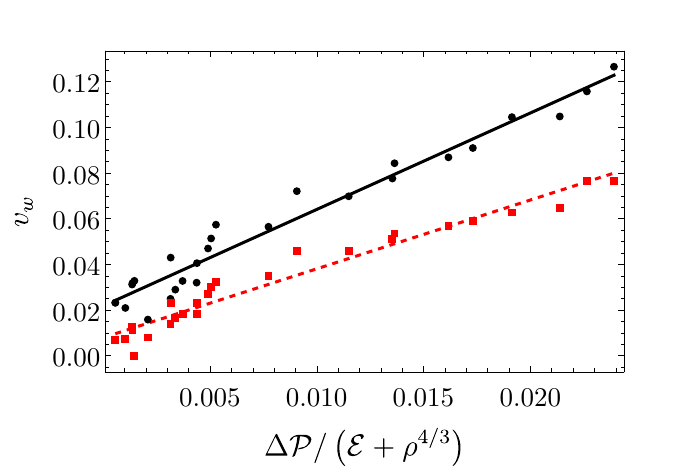}
\includegraphics[width=.48\textwidth]{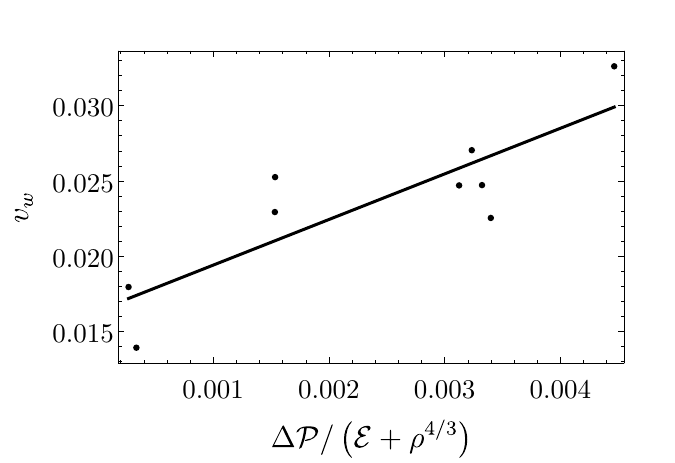}
\caption{Wall velocity as a function of the pressure difference between the inside and outside of the bubble, normalized by the outside energy density, for  supercooled (left) and  superheated (right) bubbles. The solid black lines correspond to  linear fits to  the data. In the left panel we have added red squares corresponding to the prediction based on a  large jump in the number of degrees of freedom, as well as a linear fit to them in dashed red. Both linear fits seem to have a non-zero intercept, presumably due to the lack of data at very small velocities.  
}
\label{fig:linear_fit_speed}
\end{figure}

Ref.~\cite{Bea:2021zol} identified  a phenomenological, approximately linear relationship between the wall velocity and the pressure difference between the inside and outside of the bubble, normalized by the energy density of the exterior metastable state:
\begin{equation}
        v_w\propto\frac{\Delta\mathcal{P}}{\mathcal{E}} \,.
        \label{eq:linear_speed_relation_OLD}
\end{equation}
The physical intuition behind this relation is that $\Delta P$ provides the force that tends to accelerate the bubble, whereas $\mathcal{E}$ provides the resistance to this acceleration.  
The proportionality constant in \eqref{eq:linear_speed_relation_OLD} was shown    to be calculable in terms of only equilibrium properties \cite{Janik}.   A natural generalization in the presence of the baryon density is  

\begin{equation}
        v_w\propto\frac{\Delta\mathcal{P}}{\mathcal{E} + \rho^{4/3}} \,.
        \label{eq:linear_speed_relation}
\end{equation}
We have examined the validity of \eqref{eq:linear_speed_relation_OLD} and \eqref{eq:linear_speed_relation} in the present case, obtaining almost identical results. In Fig.~\ref{fig:linear_fit_speed} we assess the validity of \eqref{eq:linear_speed_relation} for supercooled (black dots in the top panel) and superheated (right panel) bubbles. We have added solid black lines corresponding to linear fits to the data. While in the supercooled case the general trend of the data seems consistent with a linear behavior, in the superheated case this is unclear. Moreover, in the supercooled case we find a much larger dispersion around the linear fit than in \cite{Bea:2021zol}, despite the fact that the range of velocities covered here is  smaller than in that reference.

Refs.~\cite{Sanchez-Garitaonandia:2023zqz} provided approximate expressions for the wall velocity under the assumption that the jump in the number of degrees of freedom across the transition is large.  

This  is encoded in a small parameter defined as

\begin{equation}
    \delta = \frac{(\mathcal{E}+\mathcal{P})_L}
    {(\mathcal{E}+\mathcal{P})_H}\ll 1 \,,
\end{equation}
where $H$ and $L$ denote the high- and the low-density phases, respectively.  In \cite{Bea:2024bxu}, this approximation was applied to the study of superheated bubbles. In essence, it amounts to assuming that the state on the high-energy phase lies on the critical line of FOPT. For supercooled bubbles, this allows for a prediction of the wall velocity for each metastable state by integrating the hydrodynamic equations up to the  location of the bubble wall. For superheated bubbles, this approximation just leads to the conclusion that the wall velocity is largely suppressed (see \cite{Sanchez-Garitaonandia:2023zqz, Bea:2024bxu} for details).

Let us start with supercooled bubbles. Solving the hydrodynamic equations reduces to solving the shock matching conditions
\begin{equation}
    v_+v_- = \frac{\mathcal{P}_+-\mathcal{P}_-}{\mathcal{E}_+-\mathcal{E}_-},\quad \frac{v_+}{v_-} = \frac{\mathcal{E}_--\mathcal{P}_+}{\mathcal{E}_+-\mathcal{P}_-},\quad 
    \frac{\rho_+v_+}{\sqrt{1-v_+^2}} = \frac{\rho_-v_-}{\sqrt{1-v_-^2}} \,,
\end{equation}
with the assumption that the state behind the shock,  $(\mathcal{E}_-,\rho_-)$, lies on the line of FOPT, which can be parametrized as $\rho=\rho_c(\mathcal{E})$. The velocity $v_-$ corresponds to the wall velocity boosted to the shock rest frame, i.e.
\begin{equation}
    v_- = \frac{v_w-v_s}{1-v_w v_s}\,.
\end{equation}
For a given metastable state specified by values $(\mathcal{E}, \rho)$, this leads to three equations with three unknowns: $v_w$, $\mathcal{E}_-$ and $v_s$. The resulting values of the wall velocity are shown as red squares in  Fig.~\ref{fig:linear_fit_speed}(left). 
\begin{figure}[tbp]
\centering
\includegraphics[width=.455\textwidth]{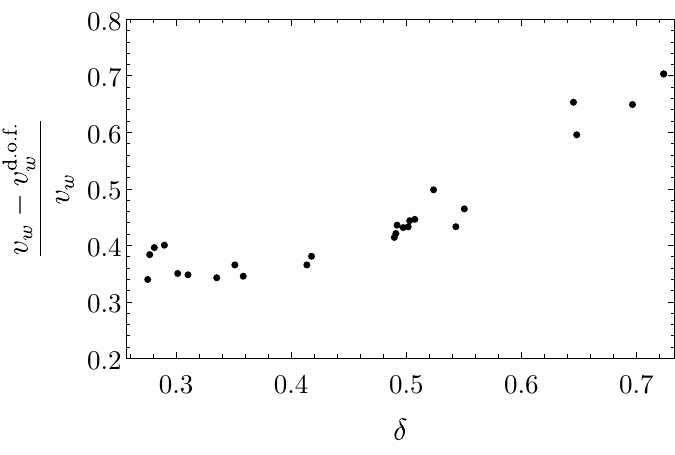}
\includegraphics[width=.47\textwidth]{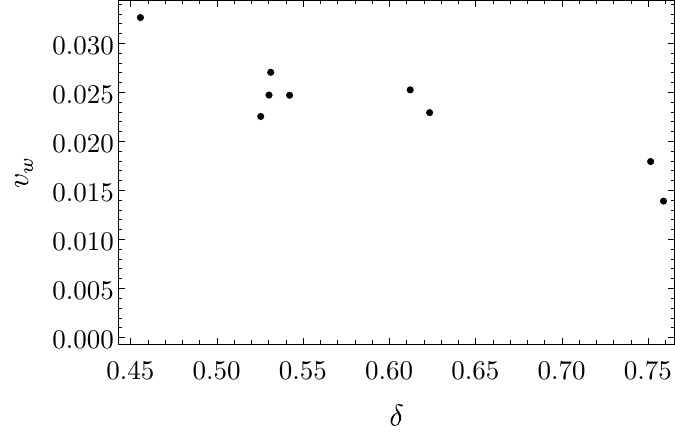}
\caption{(Left) Relative difference between the exact wall velocity, $v_w$, and the prediction based on a large number of degrees of freedom, $v_w^{\rm{d.o.f.}}$, for supercooled bubbles. (Right) Wall velocity of superheated bubbles as a function of the jump in the number of degrees of freedom. }
\label{fig:large_jump_dof}
\end{figure}
We observe that they lie below the exact values and are laid out in a very similar fashion, with comparable dispersion around a general, linear trend with lower slope (red dashed line). The discrepancy ranges from 35\% to 70\%, being smallest at the lowest value of $\delta \simeq 0.3$. In QCD, a value of $\delta \simeq 0.1$ may be expected far enough from the CP. Since the error decreases with $\delta$, this suggests that the large-jump approximation may provide a reasonably accurate estimate for the velocity of supercooled QCD bubbles.

In the case of superheated bubbles, the large-jump approximation only leads to the conclusion that $v_w\propto \delta$. The proportionality constant depends on the exact states on each side of the wall and cannot be fixed without the numerical simulations. Nevertheless, one would expect $v_w$ to decrease with $\delta$. In   
Fig.~\ref{fig:large_jump_dof}(right) we observe the opposite behavior. 
This indicates that the $\delta$ values that we have considered are too large for the approximation in Ref.~\cite{Sanchez-Garitaonandia:2023zqz} to be applicable to superheated bubbles. The same conclusion may apply to QCD, where a microscopic analysis may  be required to determine the velocity of superheated  bubbles.

\section{Discussion}
Gravitational waves from neutron star mergers may provide direct access to the conjectured line of FOPTs in the QCD phase diagram \cite{Casalderrey-Solana:2022rrn}. Maximizing the discovery potential requires a theoretical understanding of the dynamics of the transition, which proceeds through the nucleation of bubbles of the stable phase inside the metastable phase. 

We have performed the first microscopic simulation of the dynamics of these bubbles in a holographic model whose phase diagram, shown in \fig{fig:phase_diagram}, mirrors the expected QCD phase diagram at the qualitative level. In addition to the line of FOPT ending at the CP, in the figure we also show the boundaries of the metastable regions. These play a crucial role in our analysis, since they are the regions where bubbles can be nucleated. The time evolutions of a supercooled and a superheated bubble are shown in Figs.~\ref{fig:supercooled_3D} and \ref{fig:superheated_3D}, respectively. 

From our simulations we determine the dependence of the bubble wall velocity on the nucleation point. One of our conclusions is rather intuitive: the wall velocity increases monotonically with the amount of supercooling or superheating. This means that the velocity increases as one moves away from the critical line towards the boundaries of the metastable regions, as illustrated in \fig{fig:vw_vs_distance_to_turning}, and that the velocity increases with the distance to the CP, as shown in \fig{fig:vw_vs_distance_to_CP}. We have also observed that the wall velocity of superheated bubbles is generally smaller than that of supercooled bubbles for comparable amounts of superheating and supercooling. This is reflected in  the different maximum velocities \eqref{vmax} that we were able to obtain for superheated and supercooled bubbles.

These features may play an important role in determining the gravitational wave spectrum produced by the transition, since this is highly sensitive to the wall velocity. In this respect, it is crucial to assess their model (in)dependence. For this purpose, it would be interesting to apply  our analysis to other models, in particular to those that incorporate quantitative QCD information, such as~\cite{DeWolfe:2010he,Critelli:2017oub,Grefa:2021qvt,Tootle:2022pvd,Hippert:2023bel,Shah:2024img}. This future work should pay particular attention to the region of low temperature and high chemical potential, since our results indicate that the wall velocity is the highest in this part of the phase diagram. Technically, this is the most challenging region in our model, and we expect this to be the case in other models too. 
It would also be interesting to compare our results with a hydrodynamic description of the bubble dynamics. This would require extending the analysis of \cite{Bea:2024bxu} to a non-conformal EoS like the one considered here. In the hydrodynamic description, spherical bubbles can be easily studied. It would be interesting to adapt our code to obtain a microscopic description of  spherical bubbles. 

We have focused on the  wall velocity of isolated bubbles since this is a crucial parameter for the estimation of the gravitational wave signal. However, a more precise prediction requires a complete simulation in which bubble collisions and the resulting fluid flows are taken into consideration. Moreover, a complete study should also include the dynamics of the other parts of the neutron stars which, while not undergoing a phase transition, are nevertheless evolving hydrodynamically. This type of analysis  will be discussed elsewhere.

\appendix
\addtocontents{toc}{\protect\setcounter{tocdepth}{1}}

\section{Implementation in \texttt{Jecco}}\label{sec:Jecco_Details}
Here we will explain in further detail the equations that evolved in time, derived from~\eqref{eq:eoms} using ansatz~\eqref{eq:ansatz}, and the redefinitions performed for numerical convenience. The rest of the numerical implementation is not covered, but the interested reader can find an exhaustive description in \cite{Jecco}.

\subsection{Equations of motion}
Upon substitution of \eqref{eq:ansatz} in \eqref{eq:eoms}, we obtain the following system of differential equations
\begingroup
\allowdisplaybreaks
\begin{align*}
    S''+\left(\frac{1}{2}B^{\prime 2}+\frac{2}{3}\phi^{\prime 2}\right)S+\frac{fe^{2B}A_x^{\prime 2}}{3S} & = 0,\\
    \Psi_t' + \frac{\partial_{\phi}f}{f}\phi'\Psi_t + \frac{\partial_{\phi}fSe^{2B}A_x'\partial_x\phi}{f}+\partial_x\left(S e^{2B} A_x'\right) & = 0, \\
    F'' + \left(2B'+\frac{S'}{S}\right)F'+\left(2B''+B^{\prime 2}+\frac{4}{3}\phi^{\prime 2}-4\frac{S^{\prime 2}}{S^2}+6\frac{B'S'}{S}+\frac{2fe^{2B}A_x^{\prime 2}}{3S^2}\right)F - & \\
    -3B'\partial_x B-2\partial_xB' -4\phi'\partial_x\phi-\frac{4\partial_xS'}{S}-\frac{4S'\partial_xS}{S^2}-6\frac{B'\partial_xS}{S}+\frac{2f(\phi)\Psi_tA_x'}{S^3} & = 0, \\
    \dot{S}'+\frac{2S'}{S}\dot{S} + \frac{2}{3}SV +\frac{e^{2B}}{3S}\tilde{\phi}^2+\frac{e^{2B}}{3S}\tilde{\tilde{B}}+\frac{7e^{2B}}{12S}\tilde{B}^2+\frac{2e^{2B}}{3S^2}\tilde{\tilde{S}} - \frac{e^{2B}}{3S^2}\tilde{S}^2+ & \\
    +\frac{4e^{2B}}{2S^2}\tilde{B}\tilde{S}-\frac{e^{2B}}{6S^2}\left(2S\tilde{B}+\tilde{S}\right)F'-\frac{e^{2B}}{12S}F^{\prime 2}-\frac{e^{2B}}{6S}\tilde{F}'+\frac{f}{6}SA_t^{\prime 2} & = 0,\\
    \dot{\phi}' + \frac{3S'}{2S}\dot{S} - \frac{\partial_{\phi}f}A_x'{4S^2}\dot{A_x} - \frac{\partial_{\phi}V}{2}+\frac{e^{2B}}{2S^2}\tilde{\tilde{\phi}}+\frac{e^{2B}}{2S^3}\left(\tilde{S}+S(2\tilde{B}-F')\right)\tilde{\phi}+ & \\
    +\frac{3\dot{S}}{2S}\phi'+\frac{\partial_{\phi}f}{8}A_t^{\prime 2}+\frac{e^{2B}\partial_{\phi}A_x'}{4S^2}\tilde{A_t} & = 0, \\
    \dot{A_x}' + \left(B'+\frac{S'}{2S}+\frac{\partial_{\phi}f\phi'}{2f}\right)\dot{A_x}+\frac{\partial_{\phi}fA_x'}{2f}\dot{\phi} + A_x'\dot{B} - \frac{1}{2}\tilde{A_x}'- & \\
    -\frac{1}{2}\left(2B'+\frac{S'}{S}\right)\tilde{A_t}+\frac{\dot{S}}{2S}A_x'-\frac{\partial_{\phi}f\phi'}{2f}\tilde{A_t} & = 0,\\
    \dot{B}'+\frac{3S'}{2S}\dot{B} - \frac{2fe^{2B}A_x'}{3S^2}\dot{A_x} - \frac{e^{2B}}{6S^2}\tilde{\tilde{B}}-\frac{e^{2B}}{3S^3}\tilde{\tilde{S}}-\frac{e^{2B}}{6S^2}\tilde{B}^2 + &\\
    +\frac{2e^{2B}}{3S^4}\tilde{S}^2-\frac{e^{2B}}{6S^3}\tilde{B}\tilde{S}-\frac{e^{2B}}{6S^3}(4\tilde{S}-S\tilde{B})F' + & \\
    +\frac{e^{2B}}{6S^2}F^{\prime 2} 
    +\frac{e^{2B}}{3S^2}\tilde{F}'+ \frac{3}{2S}B'\dot{S}-\frac{2e^{2B}}{3S^2}\tilde{\phi}^2+\frac{2fe^{2B}}{3S^2}A_x'\tilde{A_t} & = 0,\\
    A'' - \frac{2e^{2B}}{S^2}\tilde{\tilde{B}}-\frac{4e^{2B}}{S^3}\tilde{\tilde{S}}-\frac{7e^{2B}}{2S^2}\tilde{B}^2+\frac{2e^{2B}}{S^4}\tilde{S}^2-\frac{8e^{2B}}{S^3}\tilde{S}\tilde{B}- &\\
    -3B'\dot{B}-\frac{12}{S^2}S'\dot{S}+\frac{e^{2B}}{2S^2}F^{\prime 2}-\frac{2e^{2B}}{S^2}\tilde{\phi}^2 + &\\
    +4\phi'\dot{\phi}-\frac{4V}{3} -\frac{7}{3}fA_t^{\prime 2} -\frac{2fe^{2B}}{3S^2}A_x'\tilde{A_t}+\frac{2fe^{2B}}{3S^2}A_x'\dot{A_x} &= 0,
\end{align*}\label{eq:full_eoms}
\endgroup
where we have defined
\begin{equation}
        \dot{g}  = \partial_tg + \frac{A}{2}g',\quad
    \tilde{g}  = \partial_xg -F g',\quad
    g'  = \partial_r g,\quad
    \Psi_t = S^3A_t'-Se^{2B}FA_x'.
    \label{eq:def_psi_t}
\end{equation}
Bear in mind that the second derivatives $\tilde{g}'$ and $\dot{g}'$ imply taking the radial derivative always last, i.e.
\begin{equation*}
    \tilde{g}' = \partial_r\left(\partial_xg-Fg'\right), \qquad
    \dot{g}' = \partial_r\left(\partial_t g+\frac{A}{2}g'\right)
\end{equation*}

The system has a nested structure, with several differences with respect to the neutral equations of motion \cite{Attems:2017zam, Jecco}. 
First, the equation for $S$ is non-linear, for which a relaxation algorithm is needed. 
Second, the equations for $\Psi_t$ and $F$ are coupled to each other. We need to solve them simultaneously and only then can we obtain $A_t$ from \eqref{eq:def_psi_t}. Similarly, the equations for $\dot{\phi}$, $\dot{A_x}$ and $\dot{B}$, are coupled.

Given the initial bulk data $B(t_0,r,x)$, $\phi(t_0,r,x)$ and $A_x(t_0,r,x)$ for some initial time $t_0$, we can solve the nested structure once we provide the necessary boundary conditions. For that, we require the near-boundary expansions of all the functions.

\subsection{Near boundary expansion}

The exact near-boundary expansions introduced in \eqref{eq:fall-off} are
\begin{equation}
    \begin{aligned}
        A & = r^2 + 2\xi r +\xi^2-2\partial_t\xi-\frac{2\Lambda^2}{3} +\frac{a_4}{r^2}- \frac{2}{3r^3}\left(3\xi a_4+\partial_xf_2+\Lambda\partial_t\phi_2\right)+\cdots,\\
        B &  = \frac{b_4}{r^4}+\frac{1}{r^5}\left(\frac{2}{15}\partial_xf_2+\partial_tb_4-4\xi b_4\right)+\cdots ,\\
        S & = r+\xi -\frac{\Lambda^2}{3r}+\frac{\Lambda^2\xi}{3r^2}+\frac{1}{54r^3}\left(-18\xi^2\Lambda^2+\Lambda^4-18\Lambda\phi_2\right) \\
        & \quad+        
        \frac{\Lambda}{90^4}\left(30\Lambda\xi^3-5\Lambda^3\xi+90\phi_2\xi-24\partial_t\phi_2\right)+\cdots,\\
        F & = \partial_x\xi + \frac{f_2}{r^2}+\frac{1}{r^3}\left(\frac{4}{15}\Lambda\partial_x\phi_2-2\xi f_2-\frac{8}{5}\partial_xb_4\right)+\cdots ,\\
        \phi & = \frac{\Lambda}{r}+ \frac{\xi\Lambda}{r^2}+\frac{1}{r^3}\left(\xi^2\Lambda+\phi_2\right) + \frac{1}{r^4}\left(\partial_t\phi_2-3\xi\phi_2-\xi^3\Lambda\right)+\cdots , \\
        A_t & = \frac{A_{t,2}}{r^2}+ \frac{2}{3f(0)r^3}\left[\left(3f(0)\xi+f'(0)\Lambda\right) A_{t,2}-f(0)\partial_xA_{x,2}\right] +\cdots,\\
        A_x & = \frac{A_{x,2}}{r^2}-\frac{1}{9f(0)}\left[6f'(0)\Lambda A_{x,2}+3f(0)\left(6\xi A_{x,2}+\partial_x A_{t,2}-3\partial_tA_{x,2}\right)\right]+\cdots,
\end{aligned}\label{eq:near_bdry_full_expansion}
\end{equation}
where we work in the gauge where the boundary gauge field vanishes. Meanwhile, for the dot variables we have
\begin{equation}
\begin{aligned}
    \dot{S} & = \frac{r^2}{2}+\xi r+\frac{3\xi^2-\Lambda^2}{6}+\frac{1}{36r^2}\left(18a_4-5\Lambda^4+18\Lambda\phi_2\right)+\cdots,\\
    \dot{B} & = -\frac{2b_4}{r^3}+\cdots,\\
    \dot{\phi} & = -\frac{\Lambda}{2}+\frac{1}{r^2}\left(\frac{\Lambda^3}{3}-\frac{3}{2}\phi_2\right)+\frac{1}{r^3}\left(-\frac{2}{3}\xi\Lambda^3+3\xi\phi_2-\partial_t\phi_2\right) + \cdots,\\
    \dot{A_x} & = -\frac{A_{x,2}}{r}+\cdots.
\end{aligned}\label{eq:dot_expansion}
\end{equation}
The parameter $\xi(t,x)$, parametrizes the residual gauge transformation of \eqref{eq:ansatz}
\begin{equation}
\begin{aligned}
    r & \rightarrow r+\xi(t,x),\\
    S & \rightarrow S,\\
    B & \rightarrow B,\\
    A & \rightarrow A+2\partial_t\xi(t,x),\\
    F & \rightarrow F-\partial_x\xi(t,x),\\
    A_t & \rightarrow A_t,\\
    A_x & \rightarrow A_x,
\end{aligned}
\end{equation}
where $A_{\mu}$ does not change as we exploit the $U(1)$-gauge transformation to cancel the variation coming from the diffeomorphism $r\rightarrow r+\xi$.
The parameters present in the near-boundary expansion are not entirely independent of each other and are constraint by
\begin{equation}
    \begin{aligned}
        \partial_t a_4 & = -\frac{4}{3}\left(\partial_xf_2+\Lambda\partial_t\phi_2\right),\\
        \partial_t f_2 & = \frac{1}{4}\left(-\partial_xa_4-8\partial_xb_4+\frac{4}{3}\Lambda\partial_z\phi_2\right).\\
        \partial_t A_{t,2} & = \partial_x A_{x,2}.\\
    \end{aligned}\label{eq:boundary_constraints}
\end{equation}
Knowing the initial data $B(t_0, r, x)$, $\phi(t_0, r, x)$ and $A_x(t_0, r, x)$ allows us to read-off $b_4$, $\phi_2$, $A_{x,2}$, $\partial_t\phi_2$. Hence, providing with some initial $A_{t,2}(t_0,x)$, $f_2(t0,x)$ and $a_4(t_0,x)$ completes all the required data. Solving \eqref{eq:full_eoms} and \eqref{eq:boundary_constraints} iteratively allows us to time evolve the bulk and the boundary conditions simultaneously.

Notice that the constraints \eqref{eq:boundary_constraints} correspond to the conservation of the stress tensor, $T_{\mu\nu}$, and the $U(1)$-current, $j^{\mu}$, in \eqref{eq:conservation_laws}.

\subsubsection{Variable redefinition}

As detailed in \cite{Jecco}, we find it numerically convenient to work with two grids: one near the boundary, $g1$, and another one deeper in the bulk, $g2$. The goal is to redefine variables in $g1$ so that all the near-boundary divergent behavior is explicitly taken out of our variables, ensuring that we always work with finite quantities. The variable redefinitions in $g1$ are
\begin{equation}
    \begin{aligned}
        A &  = r^2 + 2\xi r + \left(-\frac{2\Lambda^2}{3}+\xi^2-2\partial_t\xi\right)+\frac{A_{g1}(t,r,x)}{r^2},&\\
        B &  = r^4B_{g1}(t,r,x),\\
        S & = r+\xi -\frac{\Lambda^2}{3r}+\frac{\Lambda^2\xi}{3r^2}+\frac{S_{g1}(t,r,x)}{r^3},\\
        F & = \partial_x\xi + \frac{F_{g1}(t,r,x)}{r^2},\\
        \phi & = \frac{\Lambda}{r}+ \frac{\xi\Lambda}{r^2}+\frac{\Lambda^3}{r^3}\phi_{g1}(t,r,x),\\
        A_t & = A_{t,0} +  \frac{A_{t,g1}(t,r,x)}{r^2},\\
        A_x & = A_{x,0} +  \frac{A_{x,g1}(t,r,x)}{r^2},\\
        \dot{S} & = \frac{r^2}{2}+\xi r+\frac{3\xi^2-\Lambda^2}{6}+\frac{\dot{S_{g1}}(t,r,x)}{r^2},\\
        \dot{B} & = \frac{\dot{B_{g1}}(t,r,x)}{r^3},\\
        \dot{\phi} & = -\frac{\Lambda}{2}+\frac{\dot{\phi_{g1}}(t,r,x)}{r^2},\\
        \dot{A_x} & = \frac{\dot{A_{x,g1}}(t,r,x)}{r},
    \end{aligned}\label{eq:var_redef}
\end{equation}

Upon comparison of \eqref{eq:near_bdry_full_expansion} and \eqref{eq:dot_expansion} with \eqref{eq:var_redef}, one can easily obtain the boundary conditions required for the new $g1$ variables. The $g2$ variables are simply the original ones, and their boundary condition is that the overall function must be continuous and derivable along $r$. A much more detailed exploration of this matter, as well as the rest of the numerical implementation, can be found in \cite{Jecco}.

\acknowledgments
MG is grateful to the Universitat de Barcelona for hospitality. Simulations were performed using the LEONARDO supercomputer, hosted by CINECA.

The work of MG was funded by the European Union - Next Generation EU - National Recovery and Resilience Plan (NRRP) - M4C2 CN1 Spoke2 - Research Programme CN00000013 ``National Centre for HPC, Big Data and Quantum Computing'' - CUP B83C22002830001.

YB and DM are supported by the ``Center of Excellence Maria de Maeztu 2020-2023'' award to the ICCUB (CEX2019-000918-M) funded by MCIN/AEI/10.13039/501100011033. YB and DM  also acknowledge support from grants PID2019- 105614GB-C21, PID2019-105614GB-C22, PID2022-136224NB-C21, PID2022-136224NB-C22, and 2021-SGR-872. The work of YB is also funded by a ``Beatriu de Pin\'os'' postdoctoral program
under the Ministry of Research and Universities of the Government of Catalonia (2022 BP 00225) and by a Maria Zambrano postdoctoral fellowship from the University of Barcelona.

The work of MSG was partially supported by  the European Research Council (ERC) under the European Union’s Horizon 2020 research and innovation program (grant agreement No758759).

AS acknowledges financial support from Grant No.~CEX2019-000918-M funded by  MCIN/AEI/10.13039/501100011033 and the European Research
Council (ERC) under the European Union’s Horizon 2020 research and innovation programme (grant number: 101089093 / project acronym: HighTheQ). Views and opinions expressed are however those of the authors only and do not necessarily reflect those of the European Union or the European Research Council. Neither the European Union nor the granting authority can be held
responsible for them. 

MZ acknowledges financial support by the Center for Research and Development in Mathematics and Applications (CIDMA) through the Portuguese Foundation for Science and Technology (FCT -- Fundaç\~ao para a Ci\^encia e a Tecnologia) through projects: UIDB/04106/2020 (with DOI identifier \url{https://doi.org/10.54499/UIDB/04106/2020}); UIDP/04106/2020 (DOI identifier \url{https://doi.org/10.54499/UIDP/04106/2020});  PTDC/FIS-AST/3041/2020 (DOI identifier \url{http://doi.org/10.54499/PTDC/FIS-AST/3041/2020}); 2022.04560.PTDC (DOI identifier \url{https://doi.org/10.54499/2022.04560.PTDC}); and 2022.00721.CEECIND (DOI identifier \url{https://doi.org/10.54499/2022.00721.CEECIND/CP1720/CT0001}).

\bibliography{main}{}
\bibliographystyle{JHEP}

\end{document}